\documentclass[preprint,12pt,sort&compress]{elsarticle}
\graphicspath{{./Fig/}}
\usepackage{amssymb}
\usepackage{amsthm}
\usepackage{amsfonts}
\usepackage{amsmath,bm} % this for degree symbol \degree (moze i sa $^{\circ}$)
\usepackage{gensymb}	
\usepackage{graphicx}
\usepackage{dcolumn}
\usepackage{booktabs}

\usepackage{lmodern}           % originalni fontovi proÄąË‡ireni sa Ă„Ĺ¤,Ă„â€?,ÄąÄľ i dr.
\usepackage[T1]{fontenc}       % za potpuno koriÄąË‡tenje lmodern fontova
\usepackage[cp1250]{inputenc}  % omoguĂ„â€ˇuje pisanje Ă„Ĺ¤ umjesto \v{c} i sl.
\usepackage{verbatim} 	       % for section comments
\usepackage{setspace}
\usepackage{mdframed}
\usepackage{enumitem}		   % aplhabetical lists
\journal{International Journal of Engineering Science}

\usepackage[demo]{adjustbox}
\usepackage{xcolor}
\usepackage{atbegshi}% http://ctan.org/pkg/atbegshi
\AtBeginDocument{\AtBeginShipoutNext{\AtBeginShipoutDiscard}}	% Upper and this line clears first blank page (http://tex.stackexchange.com/questions/140168/how-to-remove-a-blank-page-before-the-title-page)
\PassOptionsToPackage{hyphens}{url}\usepackage{hyperref} % rješava problem dugačkih url linkova.

\begin{document}
	\pagestyle{empty} % Removes page numbers
	\begin{titlepage}
		\color[rgb]{.4,.4,1}
		\hspace{5mm}

		\bigskip
		
		\hspace{15mm}
		\begin{minipage}{10mm}
			\color[rgb]{.7,.7,1}
			\rule{1pt}{226mm}
		\end{minipage}
		\begin{minipage}{133mm}
			\vspace{10mm}        
			\color{black}
			\sffamily
			\LARGE\bfseries On thermomechanics of multilayered beams  \\[-0.3\baselineskip]   \\[-0.3\baselineskip] 
			
			\vspace{5mm}
			{\large {Preprint of the article published in \\[-0.4\baselineskip] International Journal of Engineering Science \\[-0.1\baselineskip] 155, October 2020, 103364 }} 
			
			\vspace{10mm}        
			{\large Raffaele Barretta,\\[-0.4\baselineskip] \textsc{Marko \v{C}ana\dj{}ija}, \\[-0.4\baselineskip] Francesco Marotti de Sciarra} % Author name
			
			\large
			
			\vspace{40mm}
			%		    \small{$^*$ Corresponding author: marko.canadija@riteh.hr. Tel.: +385-51-651-496}	
			\vspace{5mm}
			
			\small
			\url{https://doi.org/10.1016/j.ijengsci.2020.103364}
			
			\textcircled{c} 2020. This manuscript version is made available under the CC-BY-NC-ND 4.0 license \url{http://creativecommons.org/licenses/by-nc-nd/4.0/}
			\hspace{30mm} % or \hfill, if you want the square sticked
			\color[rgb]{.4,.4,1} %                        to the right margin
		\end{minipage}
	\end{titlepage}

\begin{frontmatter}
	\title{On thermomechanics of multilayered beams}

\author[UN]{Raffaele Barretta}
\author[URI]{Marko \v{C}ana\dj{}ija \corref{cor}}
\ead{marko.canadija@riteh.hr}
\cortext[cor]{Corresponding author. Tel.: +385-51-651-496; Fax.: +385-51-651-490}
\author[UN]{Francesco Marotti de Sciarra}

\address[UN]{Department of Structures for Engineering and Architecture,University of Naples Federico II,Via Claudio 21,80121 Naples,Italy}
\address[URI]{University of Rijeka, Faculty of Engineering, Department of Engineering Mechanics, Vukovarska 58, 51000 Rijeka, Croatia}

	\begin{abstract}
		In this paper, the mechanical behavior of multilayered small-scale beams in nonisothermal environment is investigated. Scale phenomena are modeled by means of the mathematically well-posed and experimentally consistent stress-driven integral formulation of elasticity. The present research extends the treatment in \cite{Canadija2018} confined to elastically homogeneous nano-scopic structures. It is shown that the non-locality leads to a complex coupling between axial and transverse elastic displacements. Such a size-dependent phenomenon makes the solution of the relevant nonlocal thermoelastostatic problem, governed by a system of two ordinary differential equations with ten standard boundary conditions and non-classical constitutive boundary conditions, significantly more involved with respect to treatments in literature.  Thus, a novel solution methodology, based on Laplace transforms, is proposed and illustrated by examining simple structural schemes of current applicative interest in Nanomechanics and Nanotechnology. 
	\end{abstract}
	
\end{frontmatter}

\section{Introduction, motivation and outline}
Analysis, modelling and assessment of elastic responses of small-scale structures has been a subject of great interest in the community of Engineering Science \cite{farajpour2018review} due to the need to significantly design and optimize nanocomposites \cite{karami2019thermal,malikan2020torsional,qi2016size,omari2020modeling,eyvazian2020dynamic} and smaller and smaller technological devices \cite{malikan2020dynamics,tran2018ambient,tadi2020size,basutkar2019analytical,dehkordi2017electro,natsuki2019analysis,ghayesh2020nonlinear,vaghefpour2019nonlinear}. So, to describe the deformation of advanced materials and new-generation structural systems, one needs to modify existing elasticity continuum methodologies, and non-local mechanics seems to be a promising choice to adequately capture technically important size effets \cite{Romano17b, Romano2017c, numanouglu2018dynamic, lurie2018revisiting, lurie2019anti,pinnola2020random} in comparison with computationally expensive atomistic strategies \cite{Canadija17}. 

Like in the case of macroscopical structures, a combination of different materials in a nano- or micro-structure can lead toward superior mechanical behavior \cite{hassanzadeh2018micromechanical,xia2019tailoring,eremeyev2020flexoelectricity}. Composite beams can offer more possibilities than homogeneous structures \cite{BRACH201728,wentzel2017dependence,trofimov2018bounds,govorov2018electrical,trofimov2018inverse,liu2019nonlinear}. This has a price - more complex models are needed to describe such problems which are even more harder to solve if nonisothermal environmental conditions are involved \cite{Canadija16b, Canadija2018, taati2018buckling,dehrouyeh2019nonlinear}.

Composite beams are a potential candidate for a wide variety of nanoscopic systems. In the present research, we set a focus on 1D continua made of several layers of different materials - multilayered beams. The literature mostly deals with a special class of such structures that are known as the sandwich beams \cite{khakalo2018modelling}. A sandwich beam can be described by a central core layer, mostly thicker than two layers placed at the top and bottom surface. The thinner layers are often referred to as skins and the thickness is usually selected so the cross-section is symmetric. 

Sandwich beams operate in many engineering applications. For instance, piezoelectric effects can be involved, so these are also included in the formulation \cite{arefi2016simplified, nazemizadeh2015size, ebrahimi2018nonlocal}. A combination of sandwich and functionally graded (FG) materials including thermal effects are also considered and stability issues in the presence of dynamic effects are analyzed in \cite{al2018buckling, al2018dynamic}. For the influence of thermal phenomena in wave propagation analysis in FG sandwich beams, see \cite{arani2017influence}. Nonlinear vibration in the case of sandwich beams that are graded with carbon nanotubes is researched in \cite{arefi2017application}. Free vibration of curved sandwich-FG beams, again with various patterns of carbon nanotube distribution are examined in \cite{arefi2018free}, piezoelectric and electrostatic actuation of laminated microarches in \cite{NIKPOURIAN2019} and influence of a viscoelastic core in \cite{sobhy2019dynamic}. As a special case, formulations developed for vibrations of curved sandwich beams can be applied to straight beams by considering a very large curvature radius \cite{rahmani2016frequency}. 

Apart from the gradient-based methods deriving from Eringen-type strategies \cite{Eringen83} used in the aforementioned formulations, other methodologies are also resorted to. For instance, Joule heating in three-layered nanobeams was analyzed by modified couple stress theory in \cite{tan2019size}. Mesh-free methods can be applied to nanocomposite sandwich plates under thermomechanical loads as well, see \cite{SAFAEI201944} for an illustration in the case of two-dimensional structures.

It is known that different properties of each particular layer lead towards the shift of neutral surface \cite{arefi2019free}. The effect exists in FG structures as well, see \cite{Dehrouyeh2017} for beams and initial contributions \cite{Zhang2008,Morimoto2006,Abrate2008} for plates. Simplifications involving a symmetric structure of sandwich beam layers \cite{nazemizadeh2015size} conveniently avoid complications related to the introduction of the neutral surface shift, as it will be shown later. The same simplification benefits are obtained in FG beams graded only along axial direction \cite{rezaiee2020size}. Another approach is to simply disregard the shift in asymmetrical problems. For instance, see \cite{wu2020nonlinear,kammoun2019thermo} for research regarding nonlinear dynamics including temperature effects, dynamics influenced by imperfections in FG beams \cite{liu2019nonlinear} and \cite{kammoun2017vibration} for a comparison of Bernoulli-Euler and Timoshenko beam vibrations in three-layered beams of different thicknesses. Large differences in layers' thicknesses are also exploited for further simplifications \cite{rahmani2014frequency} since some terms in the formulation can be disregraded as small.

The research at hand does not resort to such simplifications thus aiming at a more general and accurate solution. Two main novel aspects are involved. First of all, an existing isothermal model of non-local multilayered beams \cite{canadija2020symmetry} is extended toward the nonisothermal regime. In contrast to gradient-based models described above, the present approach relies on the nonlocal stress-driven integral formulation developed in \cite{Romano17, Romano17b, Romano2017c} and extended to FG nanobeams in \cite{APUZZO2019667, Barretta2018b}. The formulation circumvents all known paradoxes and difficulties of non-local mechanics. Secondly, the presence of coupling between axial and transverse problems makes the solution of the underlying differential equations system much more complex and difficult to solve. As a non-standard approach to this issue in the absence of dynamical effects, Laplace transforms are exploited. 

The outline of the paper is as follows. In the first part, preliminary considerations about geometric and material assumptions are given. Bernoulli-Euler kinematics for the nonisothermal regime follows, after which the non-local constitutive model is presented. Starting from the equilibrium conditions, the strict variational methodology is applied to formulate the governing differential equations accompanied by suitable sets of classical and constitutive boundary conditions. Solution procedures are described in the subsequent section. Examples section demonstrates the model behavior and provides a comprehensive analysis of the developed case-studies. Main findings and implications are summarized in the concluding section.

\section{Preliminaries}
\subsection{Geometry, materials and coordinate system}
This research considers multilayered beams, that is beams consisting of $n$ layers. The model will accommodate beams in which each layer is made of a different material and has a rectangular cross-section of different dimensions. The beam has length $L$ and it is assumed that no spatial changes take place regarding cross-section and material along the beam axis. Height and width of each layer will be denoted by $h_i$ and $b_i$, $i\in \lbrace 1, 2, ..., n\rbrace$, respectively. Only cross-sections symmetric with respect to one axis will be considered. Each layer is assumed to be composed of elastic material with Young's modulus $E_i$ and coefficient of thermal expansion $\alpha_i$.

The coordinate system is defined in the following manner. The longitudinal axis is denoted by $x$, so the cross-section is positioned in the $y-z$ plane. Bending is assumed to take place in the $x-z$ plane. Consequently, the coordinates of upper and lower surfaces of a layer $i$ will be related to its height as $h_i=z_i-z_{i-1}$. The bottom of the first layer will be defined by the coordinate $z_0$.

\subsection{Nonisothermal multilayered beam kinematics}
One of the central problems in beams with stiffness varying along the bending direction ($z$) is that the neutral surface does not necessarily include the cross-sectional centroid \cite{Canadija2019, Dehrouyeh2017, Larbi2013}. Therefore, the issue known as the neutral surface shift appears in the present case as well. The shift will be denoted by $\zeta_0$. Displacements along $y$ and $z$ directions are $v(x)$ and $w(x)$. Due to the plane bending assumption, $v(x)=0$. The remaining axial displacement field of a beam depends both on longitudinal and transverse coordinates. It is described by:
\begin{equation}
\label{eq:DispField}
u(x,z)=u_0(x)+\varphi(x) (z-\zeta_0)=\frac{1}{A}\int_\Omega u(x,z) \mathrm{d}A +\varphi(x) (z-\zeta_0),
\end{equation}
where the integral term represents the average axial displacement in the cross-section $x$. For a multilayered beam, the average axial displacement of a cross-section at $x$ is  $u_0(x)=\frac{1}{A} \sum_{i=1}^{n} b_i \int_{z_{i-1}}^{z_{i}}  u(x,{z}) \mathrm{d} {z}$. Note that the term involving rotation $\varphi(x) (z-\zeta_0)$ differs from the standard one used in homogenous beams $\varphi(x) z$ by the neutral shift $\zeta_0$. 

Invoking the Bernoulli-Euler hypothesis relates the first derivative of the transverse displacement $w(x)$ to the angle of cross-sectional rotation $\varphi(x)$ as:
\begin{equation}\label{eq:PhiDispRelation2}
w^{(1)}(x) = -\varphi(x).
\end{equation}
This result can be used to express strains in a layer $i$, $z \in \left[z_{i-1}, z_i \right]$ as 
\begin{equation}\label{eq:Epsilon}
\varepsilon_i (x,z)=u_0^{(1)} -w ^{(2)}(z-\zeta_0)=\varepsilon_0-w ^{(2)}(z-\zeta_0), \quad \gamma _{xz} (x,z)=0.
\end{equation}
Above, the apex $^{(n)}$ denotes $n$-th derivative with respect to the longitudinal coordinate $x$. 

The preceding formulation is generally valid in both isothermal and nonisothermal settings. Since the focus of the present model is on non-homogenous temperature fields $\Delta \theta (x,z)$, thermal effects should be introduced. To this end, additive decomposition of normal strain into mechanical part $\varepsilon_{i,\mathrm{M}}$ and  thermal part $\varepsilon_{i,\mathrm{T}}=\alpha_i \Delta \theta$ is introduced \cite{Canadija16b}:
\begin{equation}\label{eq:StrainAddSep}
\varepsilon_i(x,z)=\varepsilon_{i,\mathrm{T}}+\varepsilon_{i,\mathrm{M}}.
\end{equation}
Now, equalizing Eqs.~(\ref{eq:Epsilon}, \ref{eq:StrainAddSep}):
\begin{equation}\label{eq:StrainEqual}
\alpha_i \Delta \theta + \varepsilon_{i,\mathrm{M}} =\varepsilon_0-w ^{(2)}(z-\zeta_0),
\end{equation}
where the mechanical strain $\varepsilon_{i,\mathrm{M}}$ is obtained from a specific non-local constitutive model.

\subsection{Non-local constitutive behavior}
The constitutive model for the mechanical strain $\varepsilon_{i,\mathrm{M}}$ is now provided. It was recently shown \cite{Canadija2019} that the thermodynamic framework suitable for the non-local beam behavior relies on the Gibbs potential. For a layer $i$, this potential has the form:
\begin{equation}\label{eq:GibbsNonLoc}
\begin{array}{c}
\rho_i g_i(\sigma_i,\Delta \theta)=-\frac{1}{2} \sigma_i \int_{0}^{L} \phi_{\lambda,i} ({x-\xi)}  E_i^{-1} {\sigma_i}(\xi,z) \mathrm{d} \xi -  \sigma_i \alpha_i \Delta \theta,
\end{array}
\end{equation}
where the kernel function $\phi_{\lambda,i} ({x)}$ is
\begin{equation}\label{eq:Kernel}
\phi_{\lambda,i}(x) = \frac{1}{2 L_{\lambda,i}} \exp \left(-\frac{\left| x\right| }{ L_{\lambda,i}}\right),
\end{equation}
${\sigma_i}={\sigma_i}(x,z)$ is the normal stress and  $L_\lambda= \lambda L$ is the characteristic length. Obviously, the characteristic length is defined in bounded domains of length $L$ multiplied by the non-dimensional small-size parameter $\lambda$. An overview of values of $\lambda$ for carbon nanotubes is given in \cite{Canadija17}. Constitutive behavior for the normal strain is then obtained from:
\begin{equation}\label{eq:2ndLaw-31}
\varepsilon_{i}=-\partial _{\sigma_i} g_i. 
\end{equation}
Using Eq.~(\ref{eq:GibbsNonLoc}), differentiation gives:
\begin{equation}\label{eq:MotStrain}
\varepsilon_{i}(x,z)= \int_{0}^{L} \phi_{\lambda,i} ({x-\xi)}  E_i^{-1} {\sigma_i}(\xi,z) \mathrm{d} \xi + \alpha_i \Delta \theta.
\end{equation}
The first term on the right-hand side represents the mechanical part of the normal strain $\varepsilon_{i,\mathrm{M}}$, while the second one is thermal strain $\varepsilon_{i,\mathrm{T}}$. This specific one-dimensional non-local stress-driven integral model was introduced in \cite{Romano2017c,Romano17,Romano17b} for homogeneous temperature fields context and extended to non-homogenous temperature fields in \cite{Canadija2018}. 

Evaluation of the latter integral is not straightforward \cite{Polyanin98, Canadija2018}. The solution 
\begin{equation}\label{eq:MotStress2DispEpsM}
\begin{array}{l}
\sigma_i = E\left( -L_{\lambda,i}^{2}(\varepsilon_{i}^{(2)}- (\alpha_i \Delta \theta)^{(2)}) +\varepsilon_{i}-\alpha_i \Delta \theta\right) ,
\end{array}
\end{equation} 
has to accommodate the constraints at the beams' ends: 
\begin{equation}\label{eq:Stress3BC}
\begin{array}{l}
E_i\left. L_{\lambda,i} \left( \varepsilon_i^{(1)}-(\alpha_i \Delta\theta)^{(1)}\right) - E_i(\varepsilon_i-\alpha_i\Delta\theta) =0 \right|_{\text{at } (0,z) } , \\
E_i\left. L_{\lambda,i} \left( \varepsilon_i^{(1)}-(\alpha_i \Delta\theta)^{(1)}\right) + E_i(\varepsilon_i-\alpha_i\Delta\theta) =0 \right|_{\text{at } (L,z) }. \\
\end{array}
\end{equation}
These constraints are known as the constitutive boundary conditions (CBC) and are typically overlooked in gradient based formulations. However, as shown in \cite{Romano2017c,Romano17,Romano17b}, inclusion of CBC effectively solves well-known paradoxes observed in literature \cite{Challamel08,Peddieson2003}.

\section{Equilibrium and constitutive boundary conditions}
With the foundations of the problem provided in the previous section, the link between displacements and stress resultants can be set-up. The formulation relies on linking displacements $u_0(x)$ and $w(x)$ to stress resultants $N(x)$ and $M(x)$, i.e. the axial force and the bending moment, respectively. With Eqs.~(\ref{eq:MotStress2DispEpsM}, \ref{eq:Stress3BC}) being the starting point, the normal strain in replaced with displacements by enforcing Bernoulli-Euler kinematics, Eq.~(\ref{eq:Epsilon}):
\begin{equation}\label{eq:MotStress2Disp}
\begin{array}{l}
\frac{{\sigma}}{E_i}   = -L_{\lambda,i}^{2} (u_0^{(3)}-w^{(4)}(z-\zeta_0)- \alpha_i \Delta \theta)+u_0^{(1)}-w^{(2)}(z-\zeta_0)- \alpha_i \Delta \theta
\end{array}
\end{equation}
and constraints:
\begin{equation}\label{eq:Stress3_1}
\begin{array}{l}
E_i L_{\lambda,i} \left( u_0^{(2)}-w^{(3)}(z-\zeta_0) - \alpha_i \Delta \theta\right)  -E_i (u_0^{(1)}-w^{(2)} (z-\zeta_0)- \alpha_i \Delta \theta)  =\left. 0 \right|_{x=0}, \\
E_i L_{\lambda,i} \left( u_0^{(2)}-w^{(3)}(z-\zeta_0) - \alpha_i \Delta \theta\right)  +E_i (u_0^{(1)}-w^{(2)} (z-\zeta_0)- \alpha_i \Delta \theta)  =\left. 0 \right|_{x=L}. \\
\end{array}
\end{equation}

In the second step, latter equations should be rearranged by means of standard equilibrium equations. For multilayered beams these equations provide stress resultants - the axial force and bending moment as:
\begin{equation}\label{eq:NM}
\begin{array}{l}
N(x) =\sum_{i=1}^{n} b_i \int_{z_{i-1}}^{z_{i}} \sigma (x,{z}) \mathrm{d} {z}, \\
M(x) = \sum_{i=1}^{n} b_i \int_{z_{i-1}}^{z_{i}} \sigma (x,{z}) ({z}-\zeta_0) \mathrm{d} {z}.\\
\end{array}
\end{equation}
Introducing stresses as defined by Eq.~(\ref{eq:MotStress2Disp}) and with notation defined in Boxes \ref{box:GeoMat}, \ref{box:Forces} and \ref{box:Stiffness} it follows:
\begin{equation}\label{eq:Ntot}
\begin{array}{l}
N(x)=\mathbf{N}\cdot\mathbf{1}=\mathbf{N}_\mathrm{M}\cdot\mathbf{1}+\mathbf{N}_\mathrm{T}\cdot\mathbf{1},\\
\end{array}
\end{equation}
where
\begin{equation}\label{eq:Ntot2}
\begin{array}{ll}
\mathbf{N}_\mathrm{M}\cdot\mathbf{1}&= -L^2(\mathbf{A}\odot\mathbf{E}\odot\bm{\lambda}^{\odot 2}) \cdot \mathbf{1} \; u_0^{(3)} +L^2(\mathbf{S}\odot\mathbf{E}\odot\bm{\lambda}^{\odot 2})\cdot \mathbf{1} \; w^{(4)} \\
&\quad+ (\mathbf{E} \odot \mathbf{A})\cdot \mathbf{1} \; u_0^{(1)}-(\mathbf{E} \odot \mathbf{S})\cdot \mathbf{1} \; w^{(2)}\\
\mathbf{N}_\mathrm{T} \cdot \mathbf{1}&=\mathbf{b}\odot\mathbf{E}\odot\bm{\alpha} \odot \bm{\Theta}_\mathrm{N}\\
\end{array}
\end{equation}
and $\odot$ denotes the Hadamard product. If the notation in Box \ref{box:Stiffness} is used, a more compact form is obtained:
\begin{equation}\label{eq:N_NL2}
\begin{array}{l}
N(x) = -k_{EA}^{2NL} u_0^{(3)} +k_{ES}^{2NL}w ^{(4)} + k_{EA} u_0^{(1)} -k_{ES} w ^{(2)}+k_{TN}.\\
\end{array}
\end{equation}

The same procedure can be repeated for the equilibrium equation regarding the bending moment Eq.~(\ref{eq:NM})$_2$. 
\begin{equation}\label{eq:Mtot}
\begin{array}{l}
M(x)=\mathbf{M}\cdot\mathbf{1}=\mathbf{M}_\mathrm{M}\cdot\mathbf{1}+\mathbf{M}_\mathrm{T}\cdot\mathbf{1},\\
\end{array}
\end{equation}
where
\begin{equation}\label{eq:M_NL}
\begin{array}{ll}
\mathbf{M}_\mathrm{M}\cdot\mathbf{1}&=-L^2(\mathbf{S}\odot\mathbf{E}\odot\bm{\lambda}^{\odot 2})\cdot \mathbf{1}  u_0^{(3)} +L^2(\mathbf{I}\odot\mathbf{E}\odot\bm{\lambda}^{\odot 2}) \cdot \mathbf{1} w ^{(4)}  \\
&\quad +(\mathbf{E} \odot \mathbf{S})\cdot \mathbf{1} u_0^{(1)} -(\mathbf{E} \odot \mathbf{I})\cdot \mathbf{1} w ^{(2)} \\
\mathbf{M}_\mathrm{T} \cdot \mathbf{1}&=\mathbf{b}\odot\mathbf{E}\odot\bm{\alpha} \odot \bm{\Theta}_\mathrm{M}\\
\end{array}
\end{equation}
or
\begin{equation}\label{eq:M_NL2}
\begin{array}{l}
M(x) = -k_{ES}^{2NL} u_0^{(3)} +k_{EI}^{2NL}w ^{(4)} + k_{ES} u_0^{(1)} -k_{EI} w ^{(2)}+k_{TM}.\\
\end{array}
\end{equation}

A similar procedure must be applied to the constitutive boundary conditions Eq.~(\ref{eq:Stress3_1}) which become for equilibrium of axial forces:
\begin{equation}\label{eq:CBC_N}
\begin{array}{l}
L(\bm{\lambda} \odot \mathbf{E} )\cdot \left( u_0^{(2)} \mathbf{A}- w^{(3)} \mathbf{S} -(\bm{\alpha} \odot \bm{\Theta}_\mathrm{N})^{(1)} \right) - \mathbf{E} \cdot (u_0^{(1)} \mathbf{A} - w^{(2)} \mathbf{S}-\bm{\alpha} \odot \bm{\Theta}_\mathrm{N}) =\left. 0 \right|_{x=0}, \\
L(\bm{\lambda} \odot \mathbf{E} )\cdot \left( u_0^{(2)} \mathbf{A}- w^{(3)} \mathbf{S} -(\bm{\alpha} \odot \bm{\Theta}_\mathrm{N})^{(1)} \right) + \mathbf{E} \cdot (u_0^{(1)} \mathbf{A} - w^{(2)} \mathbf{S}-\bm{\alpha} \odot \bm{\Theta}_\mathrm{N}) =\left. 0 \right|_{x=L}. \\
\end{array}
\end{equation}
Likewise, for equilibrium of bending moments, the constitutive boundary conditions are:
\begin{equation}\label{eq:CBC_M}
\begin{array}{l}
L(\bm{\lambda} \odot \mathbf{E} )\cdot \left( u_0^{(2)} \mathbf{S}- w^{(3)} \mathbf{I} -(\bm{\alpha} \odot \bm{\Theta}_\mathrm{M})^{(1)} \right) - \mathbf{E} \cdot (u_0^{(1)} \mathbf{S} - w^{(2)} \mathbf{I}-\bm{\alpha} \odot \bm{\Theta}_\mathrm{M}) =\left. 0 \right|_{x=0}, \\
L(\bm{\lambda} \odot \mathbf{E} )\cdot \left( u_0^{(2)} \mathbf{S}- w^{(3)} \mathbf{I} -(\bm{\alpha} \odot \bm{\Theta}_\mathrm{M})^{(1)}\right) + \mathbf{E} \cdot (u_0^{(1)} \mathbf{S} - w^{(2)} \mathbf{I}-\bm{\alpha} \odot \bm{\Theta}_\mathrm{M}) =\left. 0 \right|_{x=L}. \\
\end{array}
\end{equation}
With the notation in Box~\ref{box:Stiffness} two sets of constitutive boundary conditions are:
\begin{equation}\label{eq:CBC_NM}
\begin{array}{l}
\left( k_{EA}^{NL} u_0^{(2)} - k_{ES}^{NL} u_0^{(2)} w^{(3)} - k_{TN}^{NL} \right) - (k_{EA} u_0^{(1)} -k_{2TN}) =\left. 0 \right|_{x=0} \\
\left( k_{EA}^{NL} u_0^{(2)} - k_{ES}^{NL} u_0^{(2)} w^{(3)} - k_{TN}^{NL} \right) + (k_{EA} u_0^{(1)} -k_{2TN}) =\left. 0 \right|_{x=L} \\
\left( k_{ES}^{NL} u_0^{(2)} - k_{EI}^{NL} w^{(3)} -k_{TM}^{NL} \right) - (-k_{EI} w^{(2)} -k_{2TM}) =\left. 0 \right|_{x=0} \\
\left( k_{ES}^{NL} u_0^{(2)} - k_{EI}^{NL} w^{(3)} -k_{TM}^{NL} \right) + (-k_{EI} w^{(2)} -k_{2TM}) =\left. 0 \right|_{x=L} \\
\end{array}
\end{equation}
This completes non-local stress-driven beam relations between displacements and stress resultants.

\mdfdefinestyle{theoremstyle}{
nobreak=true
}
\mdtheorem[style=theoremstyle]{definition}{Box}

\begin{definition}[Notation for geometry and material property vectors, $(\bullet) \in \mathbb{R}^n$,  $i \in \lbrace 1,2,\hdots n \rbrace$.]
\begin{tabular}{ll}
	{Geometry:} \\   		
	Widths of layers  $\mathbf{b}$: & $  b_i$\\
	Heights of layers  $\mathbf{h}$: & $ h_i$\\
	Cross-section areas of layers $\mathbf{A}=\mathbf{b} \odot \mathbf{h}$:  & $ A_i=b_i h_i$ \\
	First moments of area of layers $\mathbf{S}$: & $ S_i=b_i\int_{z_{i-1}}^{z_{i}} ({z}-\zeta_0)    \mathrm{d} {z}$  \\
	Second moments of area of layers $\mathbf{I}$: & $ I_i=b_i\int_{z_{i-1}}^{z_{i}} ({z}-\zeta_0)^2  \mathrm{d} {z} $ \\
	\\		
	{Material:}  \\ 		
	Youngs' moduli $\mathbf{E}$: & $E_i $ \\	
	Coefficients of thermal expansion $\bm{\alpha}$: & $\alpha_i $ \\	
	%		 Characteristic lengths $\mathbf{L}_\lambda$: & $ {L}_{\lambda i}=L {\lambda_i}$ \\
	Small-size parameters $\bm{\lambda}$: & $\lambda_i$ \\	
\end{tabular}
\label{box:GeoMat}	
\end{definition}

\begin{definition}[Loading vectors, $(\bullet) \in \mathbb{R}^n$, $i \in \lbrace 1,2,\hdots n \rbrace$]
\begin{tabular}{l}	
	Temperature integral, axial part $\bm{\Theta}_\mathrm{N}$: $\Theta_{\mathrm{N}i}=\int_{z_{i-1}}^{z_{i}} \Delta\theta(x,{z}) \mathrm{d} {z}$ \\
	Temperature integral, bending part $\bm{\Theta}_\mathrm{M}$: $\Theta_{\mathrm{M}i}=\int_{z_{i-1}}^{z_{i}} \Delta\theta(x,{z}) ({z}-\zeta_0)  \mathrm{d} {z}$ \\
	Axial force, mechanical part $\mathbf{N}_\mathrm{M}:$ \\
	$N_{\mathrm{M}i}= b_i E_i \left(-L_{\lambda,i}^{2} (u_0^{(3)}h_i-w^{(4)}\int_{z_{i-1}}^{z_{i}} (z-\zeta_0)\mathrm{d} {z})\right.$ \\ 
	$\quad \quad \left.+u_0^{(1)}h_i-w^{(2)}\int_{z_{i-1}}^{z_{i}} (z-\zeta_0) \mathrm{d} {z}\right)$ \\
	Axial force, thermal part $\mathbf{N}_\mathrm{T}:$ \\
	$N_{\mathrm{T}i}=b_i E_i \alpha_i \int_{z_{i-1}}^{z_{i}} \Delta\theta(x,{z}) \mathrm{d} {z}=b_i E_i \alpha_i \Theta_{\mathrm{N}i} $ \\
	Bending moment, mechanical part $\mathbf{M}_\mathrm{M}:$ \\
	$M_{\mathrm{M}i}= b_i E_i \left(-L_{\lambda,i}^{2} (u_0^{(3)}h_i-w^{(4)}\int_{z_{i-1}}^{z_{i}} (z-\zeta_0)\mathrm{d} {z})\right.$ \\ 
	$\quad \quad \left.+u_0^{(1)}h_i-w^{(2)}\int_{z_{i-1}}^{z_{i}} (z-\zeta_0) \mathrm{d} {z}\right)$ \\
	Bending moment, thermal part $\mathbf{M}_\mathrm{T}:$ \\
	$M_{\mathrm{T}i}=b_i E_i \alpha_i \int_{z_{i-1}}^{z_{i}} \Delta\theta(x,{z}) ({z}-\zeta_0) \mathrm{d} {z}=b_i E_i \alpha_i \Theta_{\mathrm{M}i} $ \\	    
\end{tabular}
\label{box:Forces}
\end{definition}

\begin{definition}[Notation for stiffnesses]
\begin{tabular}{l}			
	$k_{EA} = \sum_{i=1}^{n} b_i E_i \int_{z_{i-1}}^{z_{i}} \mathrm{d} {z} = \sum_{i=1}^{n} A_i E_i = (\mathbf{E} \odot \mathbf{A})\cdot \mathbf{1}=\mathbf{E} \cdot \mathbf{A}$, \\		
	$k_{ES} = \sum_{i=1}^{n} b_i E_i \int_{z_{i-1}}^{z_{i}} ({z}-\zeta_0)  \mathrm{d}{z} = (\mathbf{E} \odot \mathbf{S})\cdot \mathbf{1}=\mathbf{E} \cdot \mathbf{S}$,\\
	$k_{EI} = \sum_{i=1}^{n} b_i E_i \int_{z_{i-1}}^{z_{i}} ({z}-\zeta_0)^2  \mathrm{d} {z} =(\mathbf{E} \odot \mathbf{I})\cdot \mathbf{1}=\mathbf{E} \cdot \mathbf{I}$,  \\
	$k_{EA}^{NL} = \sum_{i=1}^{n} b_i E_i h_iL_{\lambda,i} = L (\bm{\lambda} \odot \mathbf{E} ) \cdot \mathbf{A}$ \\
	$k_{ES}^{NL} =  \sum_{i=1}^{n} b_i E_i L_{\lambda,i} \int_{z_{i-1}}^{z_{i}} ({z}-\zeta_0)  \mathrm{d} {z} = L (\bm{\lambda} \odot \mathbf{E} ) \cdot \mathbf{S}$ \\
	$k_{EI}^{NL} = \sum_{i=1}^{n} b_i E_i L_{\lambda,i} \int_{z_{i-1}}^{z_{i}} ({z}-\zeta_0)^2  \mathrm{d} {z} =L (\bm{\lambda} \odot \mathbf{E} ) \cdot \mathbf{I}$ \\
	$k_{EA}^{2NL} = \sum_{i=1}^{n} b_i E_i h_iL_{\lambda,i}^{2} = L^2 (\mathbf{A}\odot\mathbf{E}\odot \bm{\lambda}^{\odot 2}) \cdot \mathbf{1}$, \\
	$k_{ES}^{2NL} = \sum_{i=1}^{n} b_i E_i L_{\lambda,i}^{2} \int_{z_{i-1}}^{z_{i}} ({z}-\zeta_0)  \mathrm{d} {z} =L^2 (\mathbf{S}\odot\mathbf{E}\odot\bm{\lambda}^{\odot 2})\cdot \mathbf{1}$, \\
	$k_{EI}^{2NL} = \sum_{i=1}^{n} b_i E_i L_{\lambda,i}^{2} \int_{z_{i-1}}^{z_{i}} ({z}-\zeta_0)^2  \mathrm{d} {z} =L^2 (\mathbf{I}\odot\mathbf{E}\odot\bm{\lambda}^{\odot 2}) \cdot \mathbf{1}$, \\
	$k_{TN}=\sum_{i=1}^{n} b_i E_i \alpha_i \int_{z_{i-1}}^{z_{i}} \Delta\theta(x,{z}) \mathrm{d} {z}=(\mathbf{b}\odot\mathbf{E})\cdot (\bm{\alpha} \odot \bm{\Theta}_\mathrm{N})$ \\
	$k_{TM}=\sum_{i=1}^{n} b_i E_i \alpha_i \int_{z_{i-1}}^{z_{i}} \Delta\theta(x,{z}) ({z}-\zeta_0) \mathrm{d} {z}=(\mathbf{b}\odot\mathbf{E})\cdot (\bm{\alpha} \odot \bm{\Theta}_\mathrm{M})$ \\
	$k_{TN}^{NL}=L(\bm{\lambda} \odot \mathbf{E} )\cdot (\bm{\alpha} \odot \bm{\Theta}_\mathrm{N})^{(1)} $ \\
	$k_{TM}^{NL}=L(\bm{\lambda} \odot \mathbf{E} )\cdot (\bm{\alpha} \odot \bm{\Theta}_\mathrm{M})^{(1)} $ \\		
	$k_{2TN}=\mathbf{E} \cdot (\bm{\alpha} \odot \bm{\Theta}_\mathrm{N})$ \\		
	$k_{2TM}=\mathbf{E} \cdot (\bm{\alpha} \odot \bm{\Theta}_\mathrm{M})$ \\			
	where: \\
	vector of ones: $\mathbf{1}= \left\lbrace \begin{matrix} 	1 &  1 & \hdots & 1 \end{matrix} \right\rbrace ^T$ of size $n$ \\		
	$\bm{\lambda}^{\odot 2}=\bm{\lambda} \odot \bm{\lambda}$ \\	
	Symbol $\odot$ represents the Hadamard product of two vectors.\\
	Symbol $\cdot$ denotes the scalar product of two vectors as usual.        
	\\	
\end{tabular}
\label{box:Stiffness}	
\end{definition}

\section{Variational formulation}
The complete differential formulation augmented with boundary conditions is now obtained by enforcing the strict varational approach. The governing potential is defined by:
\begin{equation}\label{eq:PotentialBE}
\Pi({u_0,w})=U_i-U_e,
\end{equation}
where the internal potential $U_i$ for a beam $\mathcal{B}$ is:
\begin{equation}\label{eq:PotentialInt}
\begin{array}{c}
U_i=\int_{\mathcal{B}} \int _0 ^\varepsilon \sigma \mathrm{d}{\overline{\varepsilon}}  \mathrm{d} V. \\
\end{array}
\end{equation}
The external potential $U_e$ is:
\begin{equation}\label{eq:PotentialExt}
\begin{array}{ll}
U_e=&\int_L q_x {u_0} \mathrm{d} x+\int_L q_z {w} \mathrm{d} x + \mathcal{N}_0 {u}_{0}(0) + \mathcal{N}_L {u}_{0}(L) \\
&+ \mathcal{T}_0 w(0)+ \mathcal{T}_L w(L)- \mathcal{M}_0  w^{(1)}(0)- \mathcal{M}_L  w^{(1)}(L),   \\ 
\end{array}
\end{equation}
where $\mathcal{N}_0,\; \mathcal{N}_L$ and  $\mathcal{T}_0,\; \mathcal{T}_L$ are external axial and transverse forces at $x\in\lbrace 0,L\rbrace$, respectively. $\mathcal{M}_0,\; \mathcal{M}_L$ are external moments at same positions. $q_z(x)$ and $q_x(x)$ are distributed transverse and axial loadings, respectively, 

The internal potential should be transformed into a more suitable form involving stress resultants Eq.~(\ref{eq:NM}). To this end, Eq.~(\ref{eq:Epsilon}) relating strains and displacements is introduced into the internal potential Eq.~(\ref{eq:PotentialInt}):
\begin{equation}\label{eq:PotentialInt2}
\begin{array}{c}
U_i=\int_{L} N {u}_0^{(1)} \; \mathrm{d} x -  \int_{L} M {w}^{(2)}(z-\zeta_0) \; \mathrm{d} x.\\
\end{array}
\end{equation}
Unknown displacement fields are obtained by invoking the stationarity of the minimum potential energy:
\begin{equation}\label{eq:Solution}
\begin{array}{l}
({u_0}, {w}) = \arg \underset{{u_0}, {w}}{\inf} \Pi ({u_0}, {w}).  \\
\end{array}
\end{equation}

Application of the first stationarity condition $\delta_{{u}_0} \Pi=0$ gives:
\begin{equation}\label{eq:U_u0-1}
\delta_{{u}_0} \Pi = \int_{L} N \delta {u}_0^{(1)} \; \mathrm{d} x-\int_L q_x \delta {u_0} \mathrm{d} x -\mathcal{N}_0 \delta{u}_{0}(0) - \mathcal{N}_L \delta{u}_{0}(L)=0.
\end{equation}
Performing integration by parts and grouping terms with virtual displacement $\delta u_0$ provides:
\begin{equation}\label{eq:U_u0-8b}
N^{(1)} + q_x =0,
\end{equation}
what upon application of Eq.~(\ref{eq:N_NL2}) provides the governing differential equation:
\begin{equation}\label{eq:UW}
- k_{EA}^{2NL}  u_0^{(4)} +k_{ES}^{2NL}w ^{(5)} + k_{EA} u_0^{(2)}-k_{ES} w ^{(3)}+k_{TN}^{(1)}+ q_x =0.
\end{equation}
Boundary conditions follow from considering terms with virtual displacements at the beam's end $\delta u_0(0)$ and $\delta u_0(L)$, again by virtue of Eq.~(\ref{eq:N_NL2}):
\begin{equation}\label{eq:U_u0_4}
\begin{array}{ll}
\left. (-k_{EA}^{2NL} u_0^{(3)} +k_{ES}^{2NL}w ^{(4)} + k_{EA} u_0^{(1)} -k_{ES} w ^{(2)} +k_{TN}) \right|_{x=0} =-\mathcal{N}_0  & \quad \text{ or prescribe } u_0(0), \\
\left. (-k_{EA}^{2NL} u_0^{(3)} +k_{ES}^{2NL}w ^{(4)} + k_{EA} u_0^{(1)} -k_{ES} w ^{(2)} +k_{TN}) \right|_{x=L} = \mathcal{N}_L  & \quad \text{ or prescribe } u_0(L).
\end{array}
\end{equation}
In addition to above boundary conditions, the constitutive boundary conditions Eqs.~(\ref{eq:CBC_N}) must be also enforced. 

Transverse displacements follow from the second stationary condition $\delta_{{w}} \Pi=0$:
\begin{equation}\label{eq:U_phi-1}
\begin{array}{l}
\delta_{{w}} \Pi=-\int_{L} M  \delta {w} ^{(2)}\; \mathrm{d} x  -\int_L q_z \delta {w} \mathrm{d} x -\mathcal{T}_0 \delta w(0)- \mathcal{T}_L \delta w(L) +\mathcal{M}_0  \delta w^{(1)}(0)+ \mathcal{M}_L \delta w^{(1)}(L)=0.
\end{array}
\end{equation}
Integrating by parts twice, with Eq.~(\ref{eq:M_NL2}), and assembling terms with the virtual displacement $\delta {w}$ provides:
\begin{equation}\label{eq:U_u0-35}
\begin{array}{l}
-M^{(2)} - q_z =0, \\
\end{array}
\end{equation}
or with Eq.~(\ref{eq:M_NL2}):
\begin{equation}\label{eq:WU}
\begin{array}{l}
-k_{ES}^{2NL} u_0^{(5)} +k_{EI}^{2NL}w ^{(6)} + k_{ES} u_0^{(3)} -k_{EI} w ^{(4)}+k_{TM}^{(2)} + q_z =0. \\
\end{array}
\end{equation}
Boundary conditions follow from terms with $\delta w$ at $x\in \lbrace 0,L \rbrace$ and $\delta w^{(1)}$ at $x\in \lbrace 0,L \rbrace$:
\begin{equation}\label{eq:U_BC2a}
\begin{array}{ll}
\left. (-k_{ES}^{2NL} u_0^{(3)} +k_{EI}^{2NL}w ^{(4)} + k_{ES} u_0^{(1)} -k_{EI} w ^{(2)}+k_{TM}) \right|_{x=0} = -\mathcal{M}_0 & \quad \text{ or prescribe } w^{(1)}(0),  \\
\left. (-k_{ES}^{2NL} u_0^{(3)} +k_{EI}^{2NL}w ^{(4)} + k_{ES} u_0^{(1)} -k_{EI} w ^{(2)}+k_{TM}) \right|_{x=L} =  \mathcal{M}_L & \quad \text{ or prescribe }  w^{(1)}(L)  
\end{array}
\end{equation}
while the corresponding part for $\delta w$ at $x\in \lbrace 0,L \rbrace$ gives: 
\begin{equation}\label{eq:U_BC2b}
\begin{array}{ll}
\left. (-k_{ES}^{2NL} u_0^{(4)} +k_{EI}^{2NL}w ^{(5)} + k_{ES} u_0^{(2)} -k_{EI} w ^{(3)}+k_{TM}^{(1)}) \right|_{x=0}  =-\mathcal{T}_0  & \quad \text{ or prescribe } w(0), \\
\left. (-k_{ES}^{2NL} u_0^{(4)} +k_{EI}^{2NL}w ^{(5)} + k_{ES} u_0^{(2)} -k_{EI} w ^{(3)}+k_{TM}^{(1)}) \right|_{x=L}  = \mathcal{T}_L  & \quad \text{ or prescribe } w(L).
\end{array}
\end{equation}
Similarly to the axial displacements case, the constitutive boundary conditions Eqs.~(\ref{eq:CBC_M}) must be respected as well. Finally, the axial force and bending moment can be obtained after solving for displacements by means Eq.~(\ref{eq:N_NL2}, \ref{eq:M_NL2}).

\subsection{Position of the neutral surface and coupling}
\label{sec:neutral_surface}
To determine the position of the neutral surface $\zeta_0$ we start from the following assumptions:
\begin{itemize}
\item Due to various materials involved in the composition of the cross-section, elasticity varies along with the height of the beam. As a consequence, cross-sectional rotation as implied by Bernoulli-Euler hypotheses does not take place about points laying in the plane passing through the centroid of the cross-section. Instead, these points are shifted along the $z$ axis by the (yet) unknown coordinate shift $\zeta_0$.
\item Bending does take place in the $x-z$ plane and considered cross-sections are symmetric with respect to this plane in the geometrical and material sense.
\end{itemize}
Therefore, a transverse coordinate $\zeta_0$ in which bending stresses disappear is now sought. The bending strain is described by $-w ^{(2)}(z-\zeta_0)$, cf. Eq.(\ref{eq:Epsilon}). The bending strain is related to the stress by the non-local integral law Eq.~(\ref{eq:MotStrain})
\begin{equation}\label{eq:NeuSurface1}
-w ^{(2)}E_i(z-\zeta_0)= \int_{0}^{L} \phi_{\lambda,i} ({x-\xi)} {\sigma_{\mathrm{b}i}}(\xi,z) \mathrm{d} \xi = 0
\end{equation}
where $\sigma_{\mathrm{b}i}$ denotes the bending stress in the $i$-th layer in which the neutral surface is positioned. The integral has to vanish due to assumption of the neutral surface in the layer $i$ at $\zeta_0$. Further, the left-hand side of the equation is integrated over the cross-section:
\begin{equation}\label{eq:NeuSurface2}
0= \int_{0}^{A} -w ^{(2)}E_i(z-\zeta_0) \mathrm{d} \overline{A} \quad \Rightarrow \quad 0=\sum_{i=1}^{n} b_i E_i \int_{z_{i-1}}^{z_{i}} ({z}-\zeta_0)  \mathrm{d} {z}. 
\end{equation}
According to Box~\ref{box:Stiffness}, $k_{ES} = \sum_{i=1}^{n} b_i E_i \int_{z_{i-1}}^{z_{i}} ({z}-\zeta_0)  \mathrm{d}{z}$ $= (\mathbf{E} \odot \mathbf{S})\cdot \mathbf{1}=\mathbf{E} \cdot \mathbf{S}$, so $k_{ES} =0$. This result can be used to simplify equations obtained in previous sections, as well as to obtain the position of neutral surface. In particular:
\begin{equation}\label{eq:NeuSurface3}
\sum_{i=1}^{n} b_i E_i \int_{z_{i-1}}^{z_{i}} {z}  \mathrm{d} {z} = \sum_{i=1}^{n} b_i E_i \int_{z_{i-1}}^{z_{i}} \zeta_0  \mathrm{d} {z} , 
\end{equation}
what gives the required shift of the neutral surface:
\begin{equation}\label{eq:NeuSurface4}
%\sum_{i=1}^{n} b_i E_i \int_{z_{i-1}}^{z_{i}} \overline{z}  \mathrm{d} \overline{z} = \zeta_0 \sum_{i=1}^{n} b_i E_i h_i 
%\zeta_0=\frac{\sum_{i=1}^{n} b_i E_i \int_{z_{i-1}}^{z_{i}} \overline{z}  \mathrm{d} \overline{z}}{\sum_{i=1}^{n} b_i E_i h_i} 
%\zeta_0=\frac{\sum_{i=1}^{n} b_i E_i \int_{z_{i-1}}^{z_{i}} \overline{z}  \mathrm{d} \overline{z}}{\sum_{i=1}^{n} b_i E_i h_i} 
\zeta_0=\frac{\sum_{i=1}^{n} A_i E_i (z_{i} + z_{i+1}) }{2 k_{EA}}. 
\end{equation}
The reader is cautioned that if Young modulus is a function of temperature, then the neutral surface shift is also temperature-dependent \cite{Dehrouyeh2017}.

Aforementioned simplifications of governing equations (\ref{eq:UW}, \ref{eq:WU}), boundary conditions Eqs.~(\ref{eq:U_u0_4}, \ref{eq:U_BC2a}, \ref{eq:U_BC2b}) and constitutive boundary conditions Eqs.~(\ref{eq:CBC_N}, \ref{eq:CBC_M}) due to $k_{ES}=0$ are summarized below for the reader's convenience:
\begin{equation}\label{eq:ODE}
\begin{array}{l}
-k_{EA}^{2NL}  u_0^{(4)} +k_{ES}^{2NL}w ^{(5)} + k_{EA} u_0^{(2)}+k_{TN}^{(1)}+ q_x =0 \\
-k_{ES}^{2NL} u_0^{(5)} +k_{EI}^{2NL}w ^{(6)} -k_{EI} w ^{(4)}+k_{TM}^{(2)} + q_z =0. \\
\end{array}
\end{equation}
Boundary conditions:
\begin{equation}\label{eq:BC}
\begin{array}{ll}
\left. (-k_{EA}^{2NL} u_0^{(3)} +k_{ES}^{2NL}w ^{(4)} + k_{EA} u_0^{(1)} +k_{TN}) \right|_{x=0} =-\mathcal{N}_0  & \quad \text{ or prescribe } u_0(0) \\
\left. (-k_{EA}^{2NL} u_0^{(3)} +k_{ES}^{2NL}w ^{(4)} + k_{EA} u_0^{(1)} +k_{TN}) \right|_{x=L} = \mathcal{N}_L  & \quad \text{ or prescribe } u_0(L) \\
\left. (-k_{ES}^{2NL} u_0^{(3)} +k_{EI}^{2NL}w ^{(4)} -k_{EI} w ^{(2)}+k_{TM}) \right|_{x=0} = -\mathcal{M}_0 & \quad \text{ or prescribe } w^{(1)}(0)  \\
\left. (-k_{ES}^{2NL} u_0^{(3)} +k_{EI}^{2NL}w ^{(4)} -k_{EI} w ^{(2)}+k_{TM}) \right|_{x=L} =  \mathcal{M}_L & \quad \text{ or prescribe }  w^{(1)}(L) \\ 
\left. (-k_{ES}^{2NL} u_0^{(4)} +k_{EI}^{2NL}w ^{(5)} -k_{EI} w ^{(3)}+k_{TM}^{(1)}) \right|_{x=0}  =-\mathcal{T}_0  & \quad \text{ or prescribe } w(0) \\
\left. (-k_{ES}^{2NL} u_0^{(4)} +k_{EI}^{2NL}w ^{(5)} -k_{EI} w ^{(3)}+k_{TM}^{(1)}) \right|_{x=L}  = \mathcal{T}_L  & \quad \text{ or prescribe } w(L). \\
\end{array}
\end{equation}
Constitutive boundary conditions:
\begin{equation}\label{eq:CBC}
\begin{array}{l}
\left( k_{EA}^{NL} u_0^{(2)} - k_{ES}^{NL} w^{(3)} - k_{TN}^{NL} \right) - (k_{EA} u_0^{(1)} -k_{2TN}) =\left. 0 \right|_{x=0} \\
\left( k_{EA}^{NL} u_0^{(2)} - k_{ES}^{NL} w^{(3)} - k_{TN}^{NL} \right) + (k_{EA} u_0^{(1)} -k_{2TN}) =\left. 0 \right|_{x=L} \\
\left( k_{ES}^{NL} u_0^{(2)} - k_{EI}^{NL} w^{(3)} - k_{TM}^{NL} \right) - (-k_{EI} w^{(2)} -k_{2TM}) =\left. 0 \right|_{x=0} \\
\left( k_{ES}^{NL} u_0^{(2)} - k_{EI}^{NL} w^{(3)} - k_{TM}^{NL} \right) + (-k_{EI} w^{(2)} -k_{2TM}) =\left. 0 \right|_{x=L}. \\
\end{array}
\end{equation}
At the end, it is worth emphasizing that although $k_{ES}=0$, non-local counterparts do not vanish $k_{ES}^{NL} =  \sum_{i=1}^{n} b_i E_i L_{\lambda,i} \int_{z_{i-1}}^{z_{i}} ({z}-\zeta_0)  \mathrm{d} {z} \ne 0$ and  $k_{ES}^{2NL}=\sum_{i=1}^{n} b_i E_i L_{\lambda,i}^{2} \int_{z_{i-1}}^{z_{i}} ({z}-\zeta_0)  \mathrm{d} {z} \ne 0$ in general. Thus, coupling between axial and transverse displacements in both governing equations and boundary conditions exists. Decoupling is possible only in two special cases:
\begin{itemize}
\item Small size parameters $\lambda_i$ of all layers are equal. Since lengths of all layers in a beam are always equal to the beam's length $L$, then characteristic lengths of each layer $L_{\lambda,i}$ will be equal as well. This enables taking $L_{\lambda,i}$ and $L_{\lambda,i}^{2}$ out of summation signs and subsequent cancellation. Remaining terms are then equal to $k_{ES}$ and therefore vanish.
\item Depending on particular coordinate values, integrals $\int_{z_{i-1}}^{z_{i}} ({z}-\zeta_0)  \mathrm{d} {z}$ can be positive or negative. In a special case, positive values can balance negatives, leading to $k_{ES}^{NL}=0$ and $k_{ES}^{2NL}=0$. This can take place in cross-sections symmetric about the $x-y$ coordinate plane with respect to geometrical and material properties, although some other possibilities also exist.
\end{itemize}
To summarize, coupling significantly complicates solution of the underlying system of 2 ordinary differential equations with 10 boundary conditions. Two possible solution procedures are described in the next section.

\section{Solution of the coupled problem}
\label{sec:solution}
Solution of the coupled non-homogenous ordinary differential problem Eqs.~({\ref{eq:ODE}, \ref{eq:BC}, \ref{eq:CBC}) is not straightforward. Nowadays, automated differential equations solvers like Mathematica, Matlab or alike are typically used for such purpose. In the present case, a direct approach by means of such software does not appear to have much success. Thus, two alternative solution strategies are considered below.

\subsection{Solution by substitution}
The first approach to solution of the problem Eqs.~(\ref{eq:ODE}) with the boundary conditions Eqs.~(\ref{eq:BC}, \ref{eq:CBC}) relies on the substitution:
\begin{equation}\label{eq:substitution}
\overline{u}_0= u_0^{(2)}, \quad \overline{w}= w^{(4)}.
\end{equation}
This reduces the order of the coupled system Eqs.~(\ref{eq:ODE}) to
\begin{equation}\label{eq:ODE_subs}
\begin{array}{l}
-k_{EA}^{2NL} \overline{u}_0^{(2)} +k_{ES}^{2NL}\overline{w}^{(1)} + k_{EA} \overline{u}_0+k_{TN}^{(1)}+ q_x =0 \\
-k_{ES}^{2NL} \overline{u}_0^{(3)} +k_{EI}^{2NL}\overline{w}^{(2)} - k_{EI} \overline{w}+k_{TM}^{(2)} + q_z =0. \\
\end{array}
\end{equation}
Upon solving the above coupled problem for $\overline{u}_0$ and $\overline{w}$, required displacement fields follow from decoupled equations Eqs.~(\ref{eq:substitution}). The solution will involve ten unknown integration constants that are solved from boundary conditions Eqs.~(\ref{eq:BC}, \ref{eq:CBC}). Unfortunately, this approach seems to be suitable only for the simplest problems. 

\subsection{Solution by Laplace transforms}
As a more efficient alternative to the substitution method, Laplace transforms are exploited. The central idea of Laplace transforms is to transform a problem involving ordinary differential equations into a problem based on algebraic equations. Although Laplace transforms are principally oriented toward initial value problems, the method can be successfully applied in the statical analysis of beams as well. 

Firstly, Laplace transform is applied to the system of equations (\ref{eq:ODE}) giving: 

\begin{equation}\label{eq:Laplace}
\begin{array}{l}
P_{uu}(s) U_0(s)+P_{uw}(s) W(s)+Q_u(s)=G_u(s), \\
P_{wu}(s) U_0(s)+P_{ww}(s) W(s)  +Q_w(s)=G_w(s)
\end{array}
\end{equation}
where
\begin{equation}\label{eq:Laplace3}
\begin{array}{ll}
P_{uu}(s)=&k_{EA} s^2 -k_{EA}^{2NL} s^4\\
P_{uw}(s)=&k_{ES}^{2NL} s^5 W(s)\\
Q_u(s)=&k_{EA}^{2NL} s^3 u_{00}+k_{EA}^{2NL} s^2 {u_1}+k_{EA}^{2NL} s {u_2}+k_{EA}^{2NL} {u_3}-k_{EA} u_{00}-k_{EA} {u_1}\\
&-k_{ES}^{2NL} s^4 {w_0}-k_{ES}^{2NL} s^3 {w_1}-k_{ES}^{2NL} s^2 {w_2}-k_{ES}^{2NL} s {w_3}-k_{ES}^{2NL} {w_4}\\
P_{ww}(s)=&k_{EI} s^4 -k_{EI}^{2NL} s^6\\
P_{wu}(s)=&k_{ES}^{2NL} s^5\\
Q_w(s)=&k_{EI}^{2NL} s^5 {w_0}+k_{EI}^{2NL} s^4 {w_1}+k_{EI}^{2NL} s^3 {w_2}+k_{EI}^{2NL} s^2 {w_3}+k_{EI}^{2NL} s {w_4}+k_{EI}^{2NL} {w_5}\\
&-k_{EI} s^3 {w_0}-k_{EI} s^2 {w_1}-k_{EI} s {w_2}-k_{EI} {w_3}-k_{ES}^{2NL} s^4 {u_{00}}-k_{ES}^{2NL} s^3 {u_1}\\
&-k_{ES}^{2NL}s^2 {u_2}-k_{ES}^{2NL} s {u_3}-k_{ES}^{2NL} {u_4}\\
\end{array}
\end{equation}
%\begin{equation}\label{eq:Laplace2}
%\begin{array}{l}
%(k_{EA} s^2 -k_{EA}^{2NL} s^4) U_0(s)+k_{ES}^{2NL} s^5 W(s)+k_{EA}^{2NL} s^3 u_{00}+k_{EA}^{2NL} s^2 {u_1}+k_{EA}^{2NL} s {u_2}\\
%+k_{EA}^{2NL} {u_3}=k_{EA} u_{00}+k_{EA} {u_1}+k_{ES}^{2NL} s^4 {w_0}+k_{ES}^{2NL} s^3 {w_1}+k_{ES}^{2NL} s^2 {w_2}\\
%+k_{ES}^{2NL} s {w_3}+k_{ES}^{2NL} {w_4}+K_{TN}^{(1)}+ Q_x, \\
%(k_{EI} s^4 -k_{EI}^{2NL} s^6 )W(s)+k_{ES}^{2NL} s^5 U_0(s)+k_{EI}^{2NL} s^5 {w_0}+k_{EI}^{2NL} s^4 {w_1}+k_{EI}^{2NL} s^3 {w_2}\\
%+k_{EI}^{2NL} s^2 {w_3}+k_{EI}^{2NL} s {w_4}+k_{EI}^{2NL} {w_5}=k_{EI} s^3 {w_0}+k_{EI} s^2 {w_1}+k_{EI} s {w_2}\\
%+k_{EI} {w_3}+k_{ES}^{2NL} s^4 {u_{00}}+k_{ES}^{2NL} s^3 {u_1}+k_{ES}^{2NL}s^2 {u_2}+k_{ES}^{2NL} s {u_3}+k_{ES}^{2NL} {u_4}\\
%+K_{TM}^{(2)}+ Q_z
%\end{array}
%\end{equation}
and Laplace transforms of corresponding functions are 
\begin{equation}\label{eq:Laplace4}
\begin{array}{c}
U_0(s)=\mathcal{L}(u_0(x)), \quad W(s)=\mathcal{L}(w(x)) \\
G_u(s)=\mathcal{L}(k_{TN}^{(1)}(x)+ q_x(x)), \quad G_w(s)=\mathcal{L}(k_{TM}^{(2)}(x)+ q_z(x)).
\end{array}
\end{equation}
Constants $u_{00}=u(0), u_i=u^{(i)}(0)$ and  $w_0=w(0), w_i=w^{(i)}(0)$ denote initial conditions, in the present case displacement functions and their derivatives at the origin of the coordinate system $x=0$. These constants should be eventually determined from boundary and constitutive conditions Eqs.~(\ref{eq:BC}, \ref{eq:CBC}). However, in the coupled problem at hand there are 11 initial conditions but 10 boundary and constitutive conditions, so the system seems to be underdetermined. The issue is circumvented by replacing two initial conditions terms associated with the highest order of derivatives with a single constant $uw_{45}=k_{EI}^{2NL} {w_5}-k_{ES}^{2NL} {u_4}$, thus reducing the number of initial conditions to 10. In the decoupled case described in Sec. \ref{sec:neutral_surface} $k_{ES}^{NL}=0$ and $k_{ES}^{2NL}=0$, so $uw_{45}=k_{EI}^{2NL} {w_5}$ reducing the number of integration constants by one and the latter substitution is not necessary. 

Now, Laplace transforms $U_0(s)$ and $W(s)$ are easily obtained by the linear system of equations (\ref{eq:Laplace}). With known $U_0(s)$ and $W(s)$ displacement functions follow from inverse Laplace transforms $u_0(x)=\mathcal{L}^{-1}(U_0(s))$ and $w(x)=\mathcal{L}^{-1}(W(s))$. Although evaluation of  inverse Laplace transform is straightforward, obtained terms are rather long thus being impractical for explicit documentation at this point. Finally, application of boundary and constitutive conditions Eqs.~(\ref{eq:BC}, \ref{eq:CBC}) eliminates constants associated with initial conditions introduced above.

\section{Examples}

\subsection{Nonisothermal cantilever nanobeam in a homogeneous thermal field}
The introductory example considers a simple cantilever two-layer beam, free from external mechanical loading but situated in a homogeneous temperature field $\Delta \theta=0.1$. For present purposes, measurment units are deemed not to be important, so geometrical and material properties are assumed to be dimensionless. In that way, the unit length of the non-local beams is assumed $L=1$. Also: $b_1=b_2=1$, $h_1=h_2=1$, $E_2=2E_1=2$, $\alpha_2=2\alpha_1=0.2$.

The beam's cross-section is not symmetrical with respect to material properties. Thus, the neutral surface is shifted to the new position by $\zeta_0=1/6$. 

Laplace transform of the system of differential equations is:
\begin{equation}\label{eq:Ex1_Laplace}
\begin{array}{c}
2 s^5 W(s)-9 s^4 U_0(s)+9 s^2 u_1+300 s^2 U_0(s)+9 s u_2+9 u_3\\
=2 s^2 w_2+2 s{w_3}+300 {u_1}+2 {w_4},\\
25 s^6 W(s)-1100 s^4 W(s)+24 s^3 {u_1}+24 s^2 {u_2}+24 s{u_3}+1100 s {w_2}+24 {u_4}+1100 {w_3}\\
=24 s^5 U_0(s)+25 s^3 {w_2}+25 s^2 {w_3}+25 s{w_4}+25 {w_5}.
\end{array}
\end{equation}
Solving for $U_0(s)$ and $W(s)$ gives:
\begin{equation}\label{eq:Ex1_U0sWs}
\begin{array}{ll}
U_0(s)&=\frac{s \left(3 \left(59 s^2-3300\right) u_2+177 s u_3+2 {uw45}\right)+3 \left(59 s^4-5800	s^2+110000\right)u_1+1100 (2 w_4-9 u_3)}{3 s^2 \left(59 s^4-5800 s^2+110000\right)},\\
W(s)&=\frac{s \left(s 	\left(\left(59 s^2-5800\right) (s w_2+w_3)+59 s w_4+2400 u_2+3 {uw_{45}}\right) +100 (24u_3+1100 w_2-25 w_4)\right)-100 ({uw_{45}}-1100 w_3)}{s^4 \left(59 s^4-5800	s^2+110000\right)}.
\end{array}
\end{equation}
Peforming inverse Laplace transform gives required displacements but with unknown constants included:
\begin{equation}\label{eq:Ex1_InvLap}
\begin{array}{ll}
u_0(x)&=e^{-13.59 x} \left(e^{18.66 x} \left(0.01255 u_2+0.002478 u_3-4.686\cdot 10^{-6} uw_{45}-0.001017w_4\right) \right.\\
& +e^{5.066 x} \left(0.002448 u_2-0.0002873 u_3+1.656\cdot 10^{-6}uw_{45}-0.0002137 w_4\right)\\
& +e^{22.11 x} \left(0.002448 u_2+0.0002873 u_3+1.656\cdot 10^{-6} uw_{45}+0.0002137 w_4\right) \\
& +e^{8.523 x} \left(0.01255 u_2-0.002478 u_3-4.686\cdot 10^{-6} uw_{45}+0.001017 w_4\right)\\
&\left. +e^{13.59 x} \left(-0.03 u_2+6.061\cdot 10^{-6} uw_{45}+  u_1 x-0.03 u_3 x+0.006667 w_4 x\right)\right)\\
w(x)&=e^{-13.59 x} \left(e^{5.066 x} \left(-0.0006994 u_2+0.00008207 u_3-\left(4.731\cdot 10^{-7}\right) uw_{45}+0.00006106 w_4\right) \right.\\
&+e^{22.11 x} \left(0.0006994 u_2+0.00008207 u_3+4.731\cdot 10^{-7} uw_{45}+0.00006106 w_4\right) \\
&+e^{8.523 x} \left(0.00333 u_2-0.0006573 u_3-1.243\cdot 10^{-6} uw_{45}+0.0002699 w_4\right) \\
&+e^{18.66 x} \left(-0.00333 u_2-0.0006573 u_3+1.243\cdot 10^{-6} uw_{45}+0.0002699 w_4\right) \\
&+e^{13.59 x} \left(w_4 \left(-0.01136 x^2-0.000662\right)+u_3 \left(0.01091 x^2+0.00115\right) \right. \\
&\left. \left. +x \left(-0.0001515 uw_{45} x^2+0.1667 w_3 x^2+0.5 w_2 x+0.02182 u_2-0.00002066 uw_{45}\right)\right)\right)\\
\end{array}
\end{equation}
Constants follow from solving the nonlinear system of algebraic equations:
\begin{equation}\label{eq:Ex1_const}
\begin{array}{c}
u_1=-0.01659, u_2=0.0004134, u_3=0.002107, uw_{45}-0.5818,\\
w_2=0.007167, w_3=-0.0005289, w_4=-0.002614,
\end{array}
\end{equation}
so  final results are:
\begin{equation}\label{eq:Ex1_u0w}
\begin{array}{ll}
u_0(x)&=-0.01667 x+1.981\cdot 10^{-9} e^{-8.523 x}+3.405\cdot 10^{-8} e^{-5.066 x}\\
&+0.0000158 e^{5.066 x}+9.536\cdot 10^{-8} e^{8.523 x}-0.00001593\\
w(x)&=-2.822\cdot 10^{-14} x^3+0.003636 x^2+0.00002104 x-5.66\cdot 10^{-10} e^{-8.523x}\\
&+9.034\cdot 10^{-9} e^{-5.066 x}-4.191\cdot 10^{-6} e^{5.066 x}\\
&+2.724\cdot 10^{-8} e^{8.523 x}+4.155\cdot 10^{-6}.
\end{array}
\end{equation}

The layered structure also causes bending of the beam in local and non-local cases alike. The non-local case is illustrated in Fig.~\ref{fig:Ex2_w}. The uniform axial elongation observed in local beams is also affected and a nonlinear pattern is visible at the free end of the beam, Fig.~\ref{fig:Ex2_u0}

It is well-known that homogeneous beams subjected to a constant (or linear) temperature field do not give rise to stresses \cite{Hetnarski09,Noda03,Boley1967}. However, in the case of layered beams, the issue is somewhat more complex. To illustrate this behavior, the strain field $\varepsilon$ is given in Fig.~\ref{fig:Ex2_eps}. The strain field is smooth, without discontinuities caused by the layered structure. This is not the case with the stress field, Fig.~\ref{fig:Ex2_stress_tot}. To point out the non-local character of the problem, local and non-local parts of the stress field (Eq.~(\ref{eq:MotStress2DispEpsM})) are also shown, Figs.~\ref{fig:Ex2_stress_loc}, \ref{fig:Ex2_stress_nonloc}. Such discontinuities can be attributed to differences in material properties. The magnitude of the local part is higher than the non-local part; this obviously depends on  particular values of non-local parameters.

\begin{figure}
	\centering
	\includegraphics[scale=0.42]{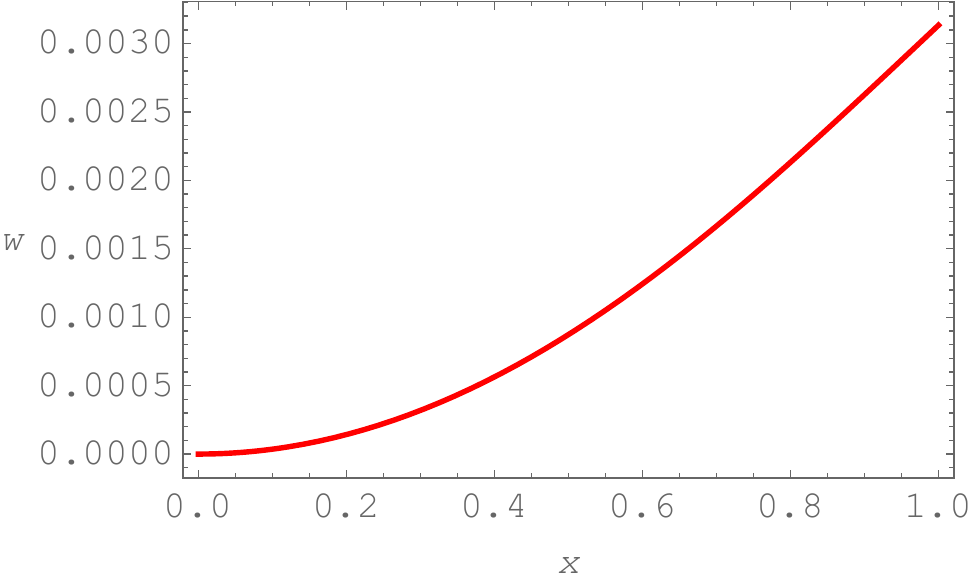}
	\caption{Transverse displacements of the beam in the homogeneous temperature field.} 
	\label{fig:Ex2_w}
\end{figure}

\begin{figure}
	\centering
	\includegraphics[scale=0.42]{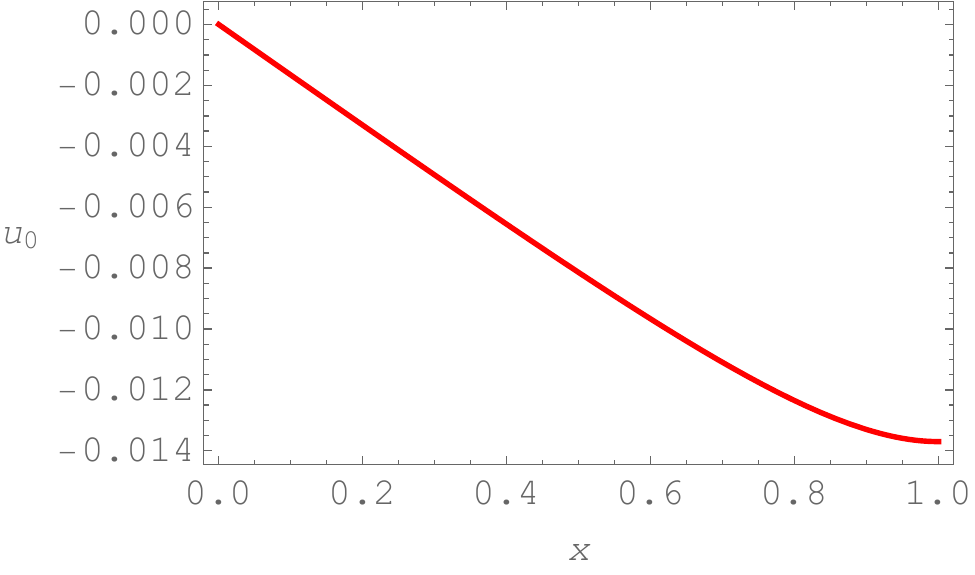}
	\caption{Axial displacements of the beam in the homogeneous temperature field.} 
	\label{fig:Ex2_u0}
\end{figure}

\begin{figure}
	\centering
	\includegraphics[scale=0.42]{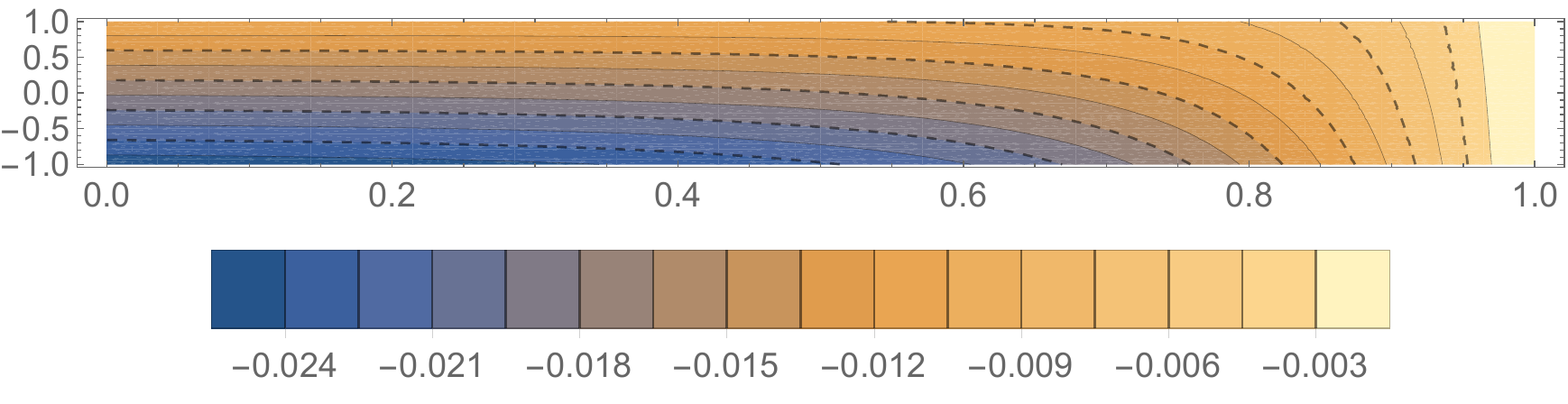}
	\caption{Normal strain distribution in the beam loaded by the homogeneous temperature field.} 
	\label{fig:Ex2_eps}
\end{figure}

\begin{figure}
	\centering
	\includegraphics[scale=0.42]{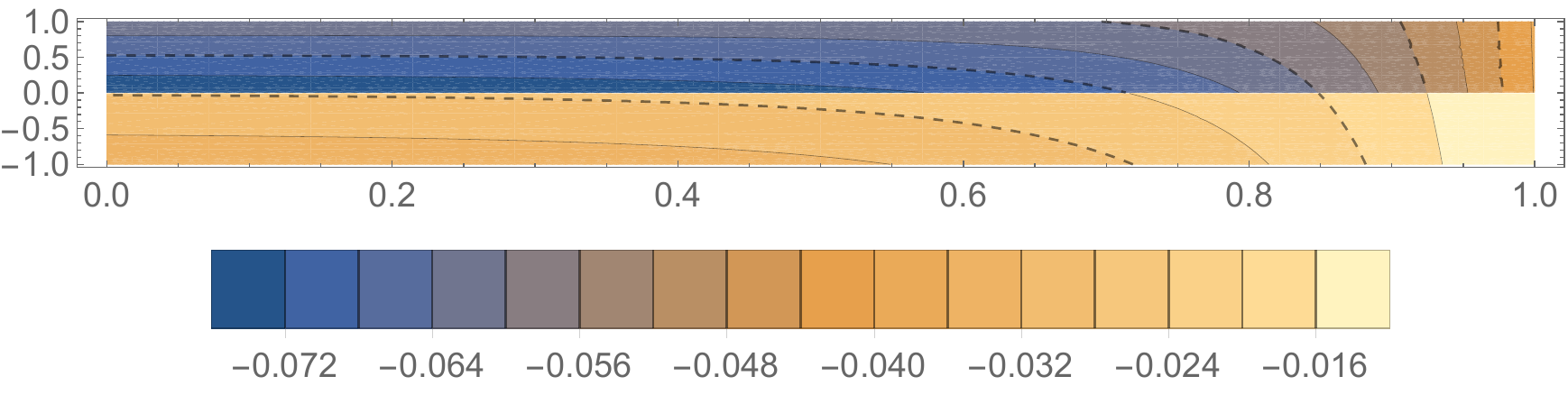}
	\caption{Local part of the stress field in the beam in the homogeneous temperature field.} 
	\label{fig:Ex2_stress_loc}
\end{figure}
\begin{figure}
	\centering
	\includegraphics[scale=0.42]{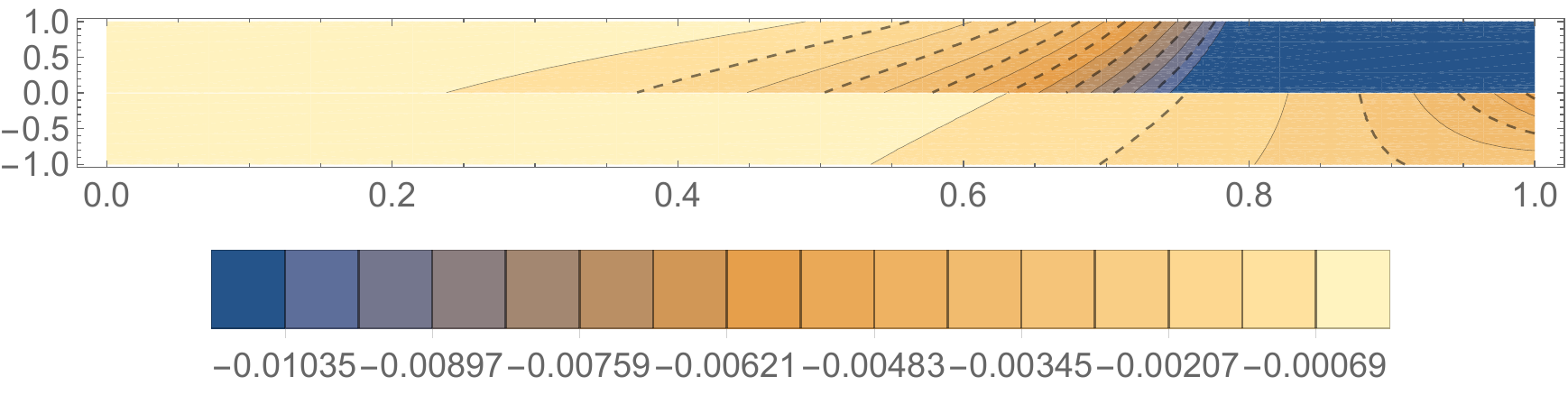}
	\caption{Non-local part of the stress field in the beam in the homogeneous temperature field.} 
	\label{fig:Ex2_stress_nonloc}
\end{figure}
\begin{figure}
	\centering
	\includegraphics[scale=0.42]{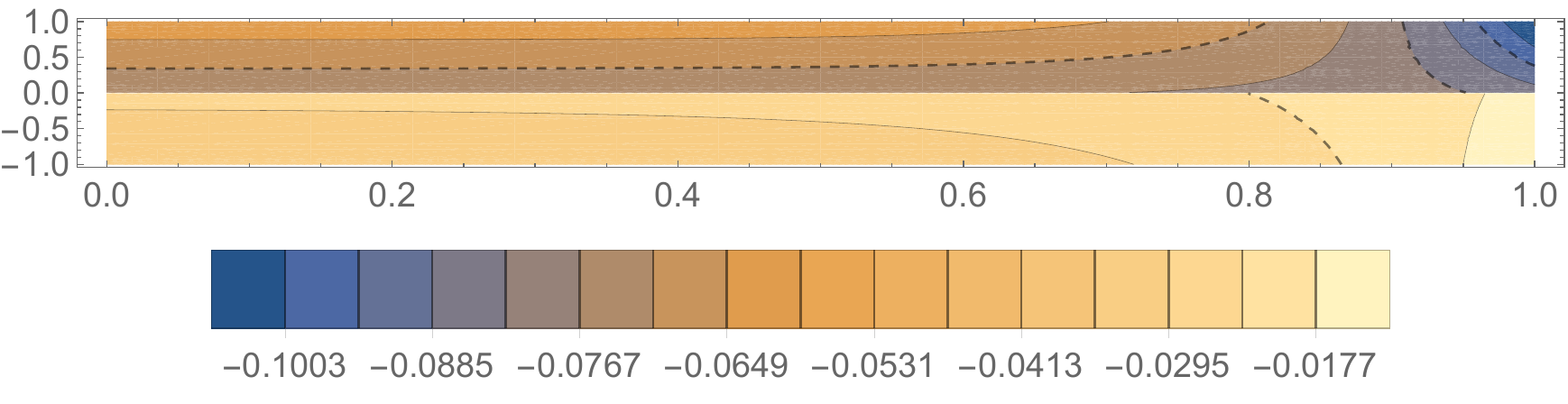}
	\caption{Total stress field (local + non-local) in the beam in the homogeneous temperature field.} 
	\label{fig:Ex2_stress_tot}
\end{figure}

\subsection{Nonisothermal cantilever piezoelectric nanobeam}
The second example considers a two-layer piezoelectric cantilever beam, Fig.~\ref{fig:Ex1_beam}. The beam is subjected to a linear temperature variation in the transverse direction $\Delta \theta (z) =0.1 z$. In addition to the temperature load, two different mechanical loads will be separately considered:
\begin{itemize}
	\item axial force at the tip $F_x$ and
	\item transverse force at the tip $F_x$.
\end{itemize}
The beam is composed of two layers: the first one is SiO$_2$ and the second one Pb(Zr$_{0.4}$Ti$_{0.6}$)O$_3$ (PZT 40/60). Material data is taken from \cite{corkovic2008development}: the Young's modulus of SiO$_2$ is $E_1=77$ GPa and the coefficient of thermal exapnsion is $\alpha_1=5.5 \times 10^{-7}$ 1/K. For the PZT 40/60 these material parameters are  $E_2=85$ GPa  and  $\alpha_2=3.7 \times 10^{-6}$ 1/K. Thicknesses of each layer are $h_1=600$ nm and $h_2=200$ nm. Width of beam's layers is $b=1200$ nm while the beam's length is $L=15000$ nm. It is necessary to determine variation of coupling terms $k_{ES}^{NL}$ and $k_{ES}^{2NL}$ with respect to non-local parameters $\lambda_1$ and $\lambda_2$. For the specific choice of $\lambda_1=0.1$ and $\lambda_2=0.2$, calculate displacement fields caused by axial force and transverse forces in the range $F_x, F_z \in \left[ 0,10000\right] $ nN.

\begin{figure}
	\centering
	\includegraphics[scale=0.42]{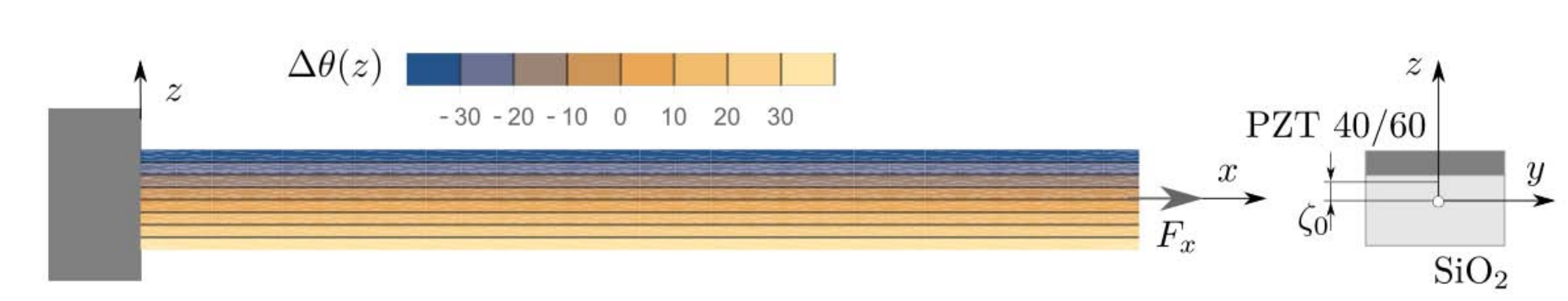}
	\caption{Cantilever beam loaded by the axial force $F_x$ and temperature field $\Delta \theta(z)$.} 
	\label{fig:Ex1_beam}
\end{figure}

Fig.~\ref{fig:Ex1_kES} presents variations in coupling terms that stem from the first moment of area. It is clearly visible that for $\lambda_1=\lambda_2$ both terms vanish, thus decoupling the problem. For all other choices of small-size parameters and given geometry and material composition, displacement fields are coupled.
\begin{figure}
	\centering
	\includegraphics[scale=0.3]{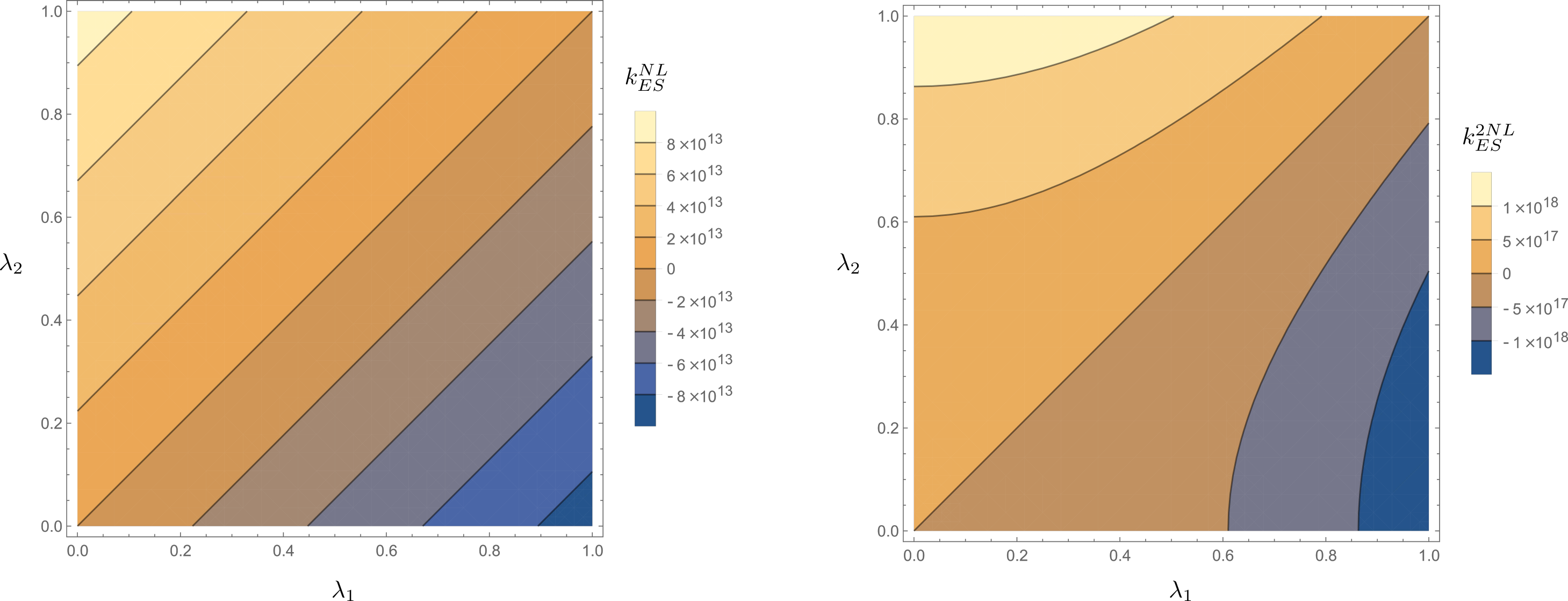}
	\caption{Variation of coupling terms $k_{ES}^{NL}$ (nN$\cdot$nm$^2$) and $k_{ES}^{2NL}$ (nN$\cdot$nm$^3$) vs. non-local parameters $\lambda_1$ and $\lambda_2$} 
	\label{fig:Ex1_kES}
\end{figure}

\subsubsection{Axial force and temperature field $\Delta \theta (z)$}
Distribution of the axial displacement and transverse displacement along the beam for different values of the axial force $F_x$ is given in Fig.~\ref{fig:Ex1_u0w}. Observe that even for the tensile axial force up to approximately 2000 nN the negative axial displacement for the complete beam is obtained. This kind of behavior is caused by coupling between axial and transverse displacements. The transverse displacement remains positive for all choices of the axial force. The influence of the axial displacement on the transverse displacement field thus remains small and the bending is caused by the temperature field.

\begin{figure}
	\centering
	\includegraphics[scale=0.3]{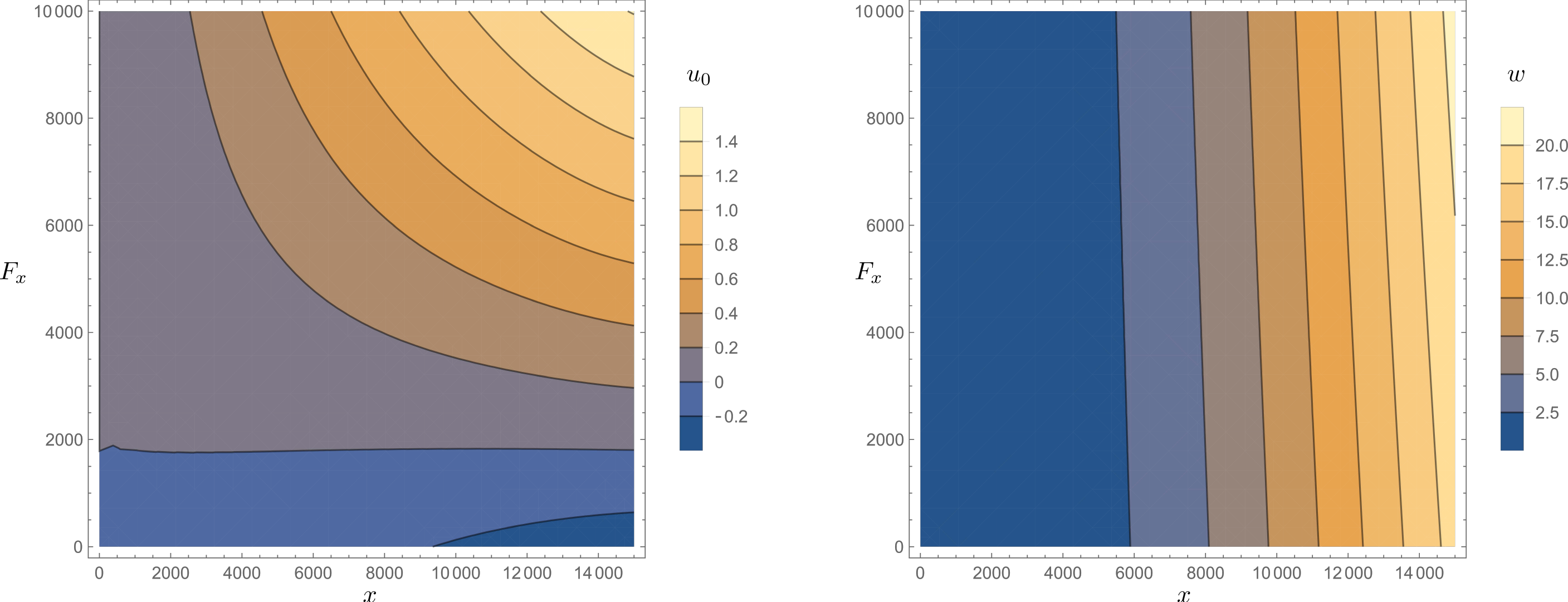}
	\caption{Variations of the axial displacement $u_0$ (nm) and transverse displacement $w$ (nm) along the beam's length $x$ (nm) vs. axial force $F_x$ (nN)} 
	\label{fig:Ex1_u0w}
\end{figure}

\subsubsection{Transverse force and temperature field $\Delta \theta (z)$}
In the case of transverse load, for the specific choice of $F_z=1000$ nN distributions of both displacements are provided in Figs.~\ref{fig:Ex1_w} and \ref{fig:Ex1_u0}. Again, Fig.~\ref{fig:Ex1_u0wFz} illustrates the variation of  axial and transverse displacements for different transverse forces and given temperature field. Bending behavior clearly dominates over much smaller axial displacements.

\begin{figure}
	\centering
	\includegraphics[scale=0.42]{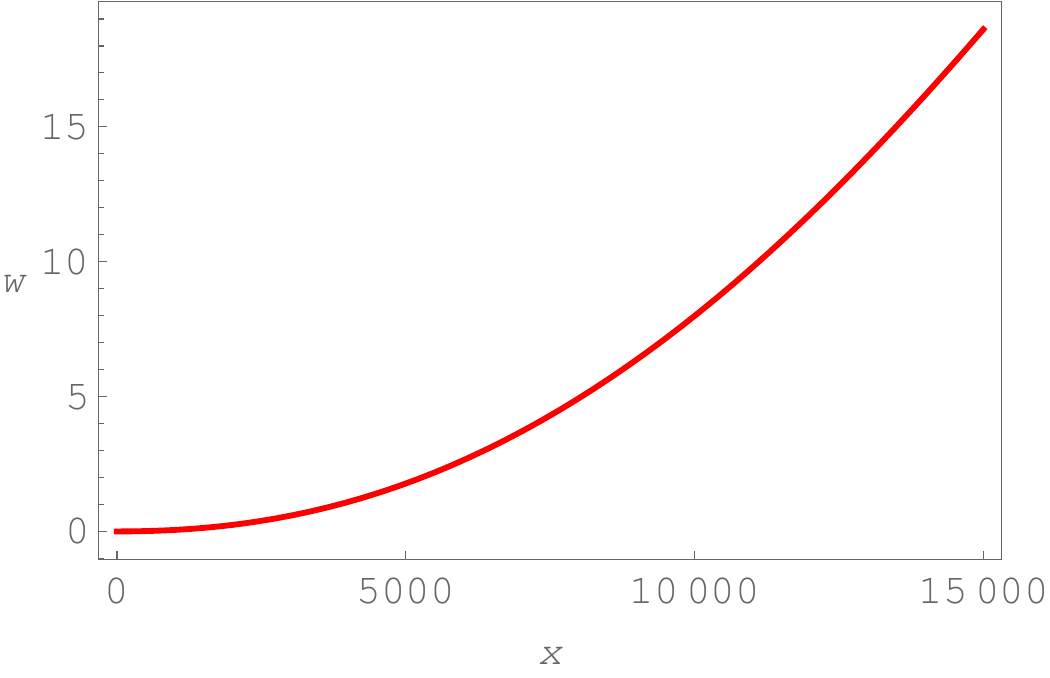}
	\caption{Transverse displacements of the beam for $F_z=1000$ nN and $\Delta \theta(z)$.} 
	\label{fig:Ex1_w}
\end{figure}

\begin{figure}
	\centering
	\includegraphics[scale=0.42]{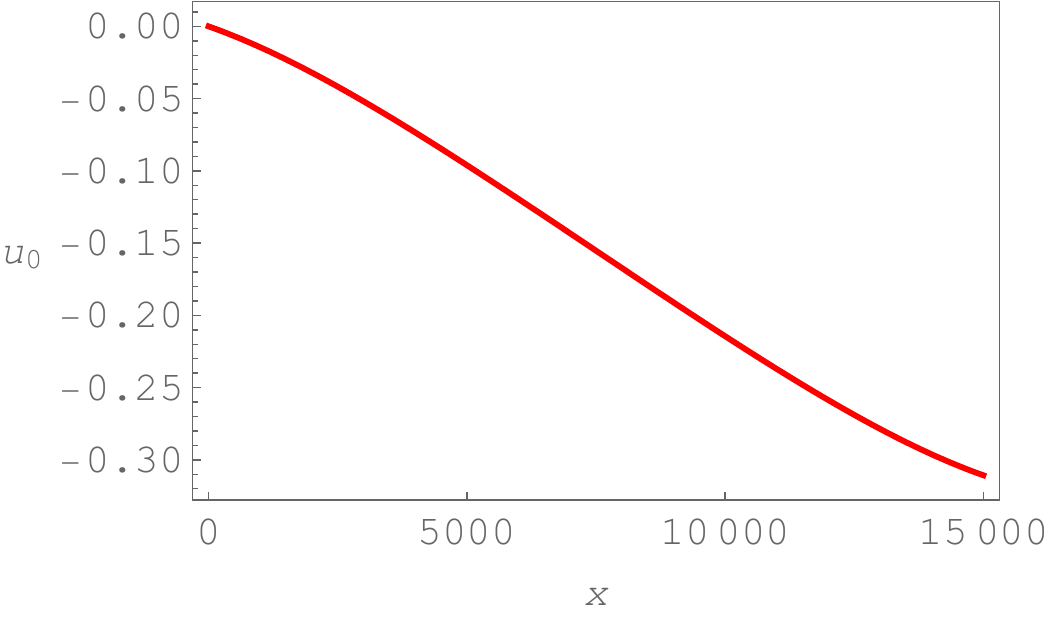}
	\caption{Axial displacements of the beam for $F_z=1000$ nN and $\Delta \theta(z)$.} 
	\label{fig:Ex1_u0}
\end{figure}

\begin{figure}
	\centering
	\includegraphics[scale=0.3]{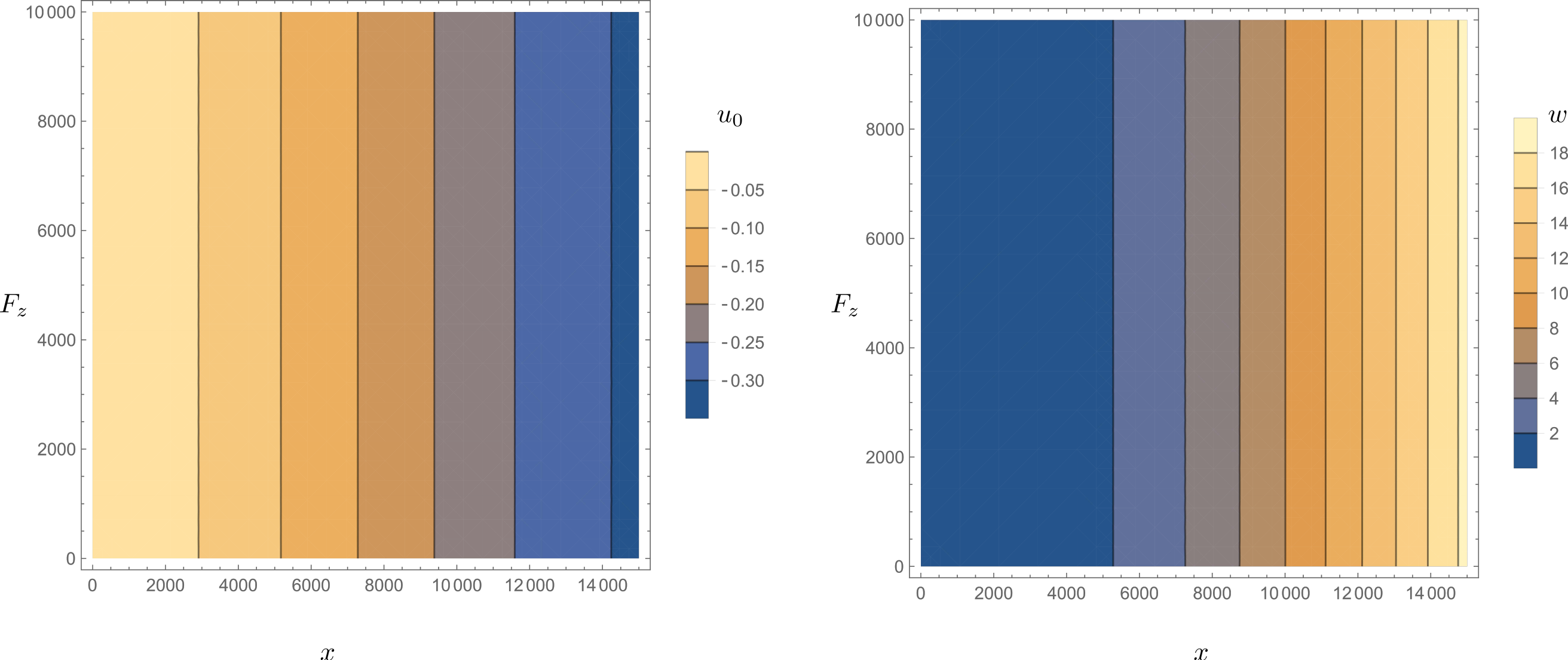}
	\caption{Variations of the axial displacement $u_0$ (nm) and transverse displacement $w$ (nm) along the beam's length $x$ (nm) vs. transverse force $F_z$ (nN)} 
	\label{fig:Ex1_u0wFz}
\end{figure}

\subsection{Thermally and mechanically loaded doubly clamped nanobeam}
The present example aims to demonstrate advantage of Laplace transforms approach in order to solve a beam loaded by nonlinear mechanical and thermal loads. Kinematically, the beam is clamped at both ends and thus statically indeterminate, Fig.~\ref{fig:Ex3_beam}. The transverse mechanical load is distributed in a parabolic manner $q_z(x)=\frac{1}{4}-(x-\frac{L}{2})^2$, while the temperature change is governed by the sinus function $\Delta \theta (x)=100 \sin (2\pi\frac{x}{L})$. The beam is assumed to be composed of two layers with following properties

\[
\begin{array}{c}
\mathbf{b}=
\begin{bmatrix}
1 \\ 
3 	
\end{bmatrix}, \quad
\mathbf{h}=
\begin{bmatrix}
3 \\ 
1 	
\end{bmatrix}, \\
\mathbf{E}=
\begin{bmatrix}
1 \\ 
2 	
\end{bmatrix}, \quad
\bm{\alpha}=
\begin{bmatrix}
0.1 \\ 
0.2 	
\end{bmatrix}, \quad
\bm{\lambda}=
\begin{bmatrix}
0.1 \\ 
0.2 	
\end{bmatrix} \\
\end{array}.
\]
Thus, the beam has T shaped cross-section. The beam is of unit length $L=1$.

This shift of the neutral surface is readiliy evaluated as $\zeta_0=1/3$. 

\begin{figure}
	\centering
	\includegraphics[scale=0.42]{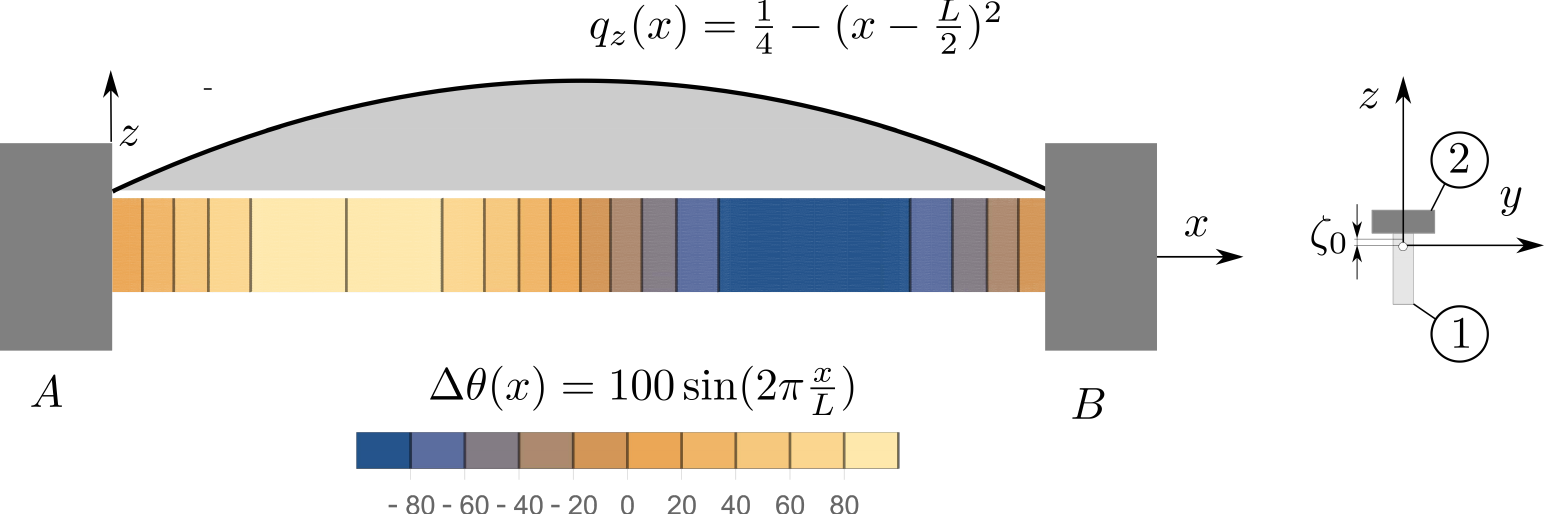}
	\caption{Doubly clamped beam loaded by the distributed parabolic load $q_z(x)$ and sinusoidal temperature field $\Delta \theta(x)$.} 
	\label{fig:Ex3_beam}
\end{figure}

The system of ordinary differential equations Eq.~(\ref{eq:ODE}) is now restated by aid of Laplace transforms in the manner of Eqs.(\ref{eq:Laplace}) as:
\begin{equation}\label{eq:Ex3_Laplace}
\begin{array}{ll}
&\frac{3}{25} s^5 W(s)-\frac{27}{100} s^4 U_0(s)-\frac{3 s^4 w_0}{25}+\frac{27 s^3 u_{00}}{100}-\frac{3 s^3 w_1}{25}+\frac{27 s^2 u_1}{100}\\
& +9 s^2 U_0(s)-\frac{3 s^2 w_2}{25}+\frac{300 \pi  s}{s^2+4 \pi ^2}+\frac{27 s u_2}{100}-\frac{3 s w_3}{25}+\frac{27 u_3}{100}-\frac{3 w_4}{25}-9 s u_{00}-9 u_1=0\\
\\
&\frac{81}{400} s^6  W(s)-\frac{3}{25} s^5 U_0(s)-\frac{81 s^5w_0}{400}+\frac{3 s^4 u_{00}}{25}-\frac{81 s^4 w_1}{400}-\frac{43}{4} s^4 W(s)\\
&+\frac{3 s^3 u_1}{25}+\frac{43 s^3 w_0}{4}-\frac{81 s^3 w_2}{400}+\frac{3 s^2 u_2}{25}+\frac{43 s^2 w_1}{4}-\frac{81 s^2 w_3}{400}+\frac{1}{s^2}+\frac{3 s u_3}{25}\\
&+\frac{43 s w_2}{4}-\frac{81 s w_4}{400}+\frac{3 u_4}{25}+\frac{43 w_3}{4}-\frac{81 w_5}{400}-\frac{2}{s^3}-\frac{320 \pi ^3}{s^2+4 \pi ^2}=0.
\end{array}
\end{equation}
Replacing two integration constants with a single one $uw_{45}=\frac{3 u_4}{25}-\frac{81 w_5}{400}$ and solving for $U_0(s), W(s)$ provides
\begin{equation}\label{eq:Ex3_Yu}
\begin{array}{ll}
U_0(s)=-\frac{1}{3 s^4 \left(s^2+4 \pi ^2\right) \left(179 s^4-21000 s^2+430000\right)} \left(-537 s^9  u_{00}-537 s^8 {u_1}-2148 \pi ^2 s^7 u_{00}\right.\\
+63000 s^7 u_{00}-537 s^7 {u_2}-2148 \pi ^2 s^6 {u_1}+63000 s^6 {u_1}-537 s^6 {u_3}+252000 \pi ^2 s^5 u_{00}\\
-1290000 s^5 u_{00}-2148 \pi ^2 s^5 {u_2}+38700 s^5 {u_2}+1600 s^5 uw_{45}-810000 \pi  s^5\\
+252000 \pi ^2 s^4 {u_1}-1290000 s^4 {u_1}-2148 \pi ^2 s^4 {u_3}+38700 s^4 {u_3}-17200 s^4 {w_4}\\
-5160000 \pi ^2 s^3  u_{00}	+154800 \pi ^2 s^3 {u_2}+6400 \pi ^2 s^3 uw_{45}-512000 \pi ^3 s^3\\
+43000000 \pi  s^3+1600 s^3	-5160000 \pi ^2 s^2 {u_1}+154800 \pi ^2 s^2 {u_3}-68800 \pi ^2 s^2 {w_4}\\
\left. -3200 s^2+6400 \pi ^2 s-12800 \pi ^2 \right) \\
\end{array}
\end{equation}
and 
\begin{equation}\label{eq:Ex3_Yw}
\begin{array}{ll}

W(s)=-\frac{1}{s^7 \left(s^2+4 \pi ^2\right) \left(179 s^4-21000 s^2+430000\right)} \left( -179 s^{12} {w_0}-179 s^{11} {w_1}-716 \pi ^2 s^{10} {w_0} \right. \\
+21000 s^{10} {w_0}-179 s^{10} {w_2}-716 \pi ^2 s^9 {w_1}+21000 s^9 {w_1}-179 s^9 {w_3}\\
+84000 \pi ^2 s^8 {w_0}-430000 s^8 {w_0}-716 \pi ^2 s^8 {w_2}+21000 s^8 {w_2}-179 s^8 {w_4}\\
-4800 s^7 {u_2}+1200 s^7 uw_{45}+84000 \pi ^2 s^7 {w_1}-430000 s^7 {w_1}-716 \pi ^2 s^7 {w_3} \\
+21000 s^7 {w_3}-160000 \pi  s^7-4800 s^6 {u_3}-1720000 \pi ^2 s^6 {w_0}+84000 \pi ^2 s^6 {w_2} \\
-430000 s^6 {w_2}-716 \pi ^2 s^6 {w_4}+8100 s^6 {w_4}-19200 \pi ^2 s^5 {u_2}+4800 \pi ^2 s^5 uw_{45}\\
-40000 s^5 uw_{45}-1720000 \pi ^2 s^5 {w_1}+84000 \pi ^2 s^5 {w_3}-430000 s^5 {w_3}\\
-384000 \pi ^3 s^5+1200 s^5-19200 \pi ^2 s^4 {u_3}-1720000 \pi ^2 s^4 {w_2}+32400 \pi ^2 s^4 {w_4}\\
-2400 s^4-160000 \pi ^2 s^3 {u_w45}-1720000 \pi ^2 s^3 {w_3}+12800000 \pi ^3 s^3+4800 \pi ^2 s^3\\
\left.-40000 s^3-9600 \pi ^2 s^2+80000 s^2-160000 \pi ^2 s+320000 \pi ^2\right).\\
\end{array}
\end{equation}
Finally, inverse Laplace transforms $\mathcal{L}^{-1}(U_0(s))$ and $\mathcal{L}^{-1}(W(s))$ give the required displacement functions that include unknown constants and support reactions. These are determined from boundary conditions Eq.~(\ref{eq:BC}) and constitutive boundary conditions Eq.~(\ref{eq:CBC}). In the present case there is $u_0(0)=u_0(L)=w(0)=w(L)=0$, $w^{(1)}(0)=w^{(1)}(L)=0$ and:
\begin{equation}\label{eq:Ex3_BC}
\begin{array}{ll}
\left. (-k_{EA}^{2NL} u_0^{(3)} +k_{ES}^{2NL}w ^{(4)} + k_{EA} u_0^{(1)} +k_{TN}) \right|_{x=L} = \mathcal{N}_L   \\
\left. (-k_{ES}^{2NL} u_0^{(3)} +k_{EI}^{2NL}w ^{(4)} -k_{EI} w ^{(2)}+k_{TM}) \right|_{x=L} =  \mathcal{M}_L  \\ 
\left. (-k_{ES}^{2NL} u_0^{(4)} +k_{EI}^{2NL}w ^{(5)} -k_{EI} w ^{(3)}+k_{TM}^{(1)}) \right|_{x=L}  = \mathcal{T}_L  \\
\left( k_{EA}^{NL} u_0^{(2)} - k_{ES}^{NL} w^{(3)} - k_{TN}^{NL} \right) - (k_{EA} u_0^{(1)} -k_{2TN}) =\left. 0 \right|_{x=0} \\
\left( k_{EA}^{NL} u_0^{(2)} - k_{ES}^{NL} w^{(3)} - k_{TN}^{NL} \right) + (k_{EA} u_0^{(1)} -k_{2TN}) =\left. 0 \right|_{x=L} \\
\left( k_{ES}^{NL} u_0^{(2)} - k_{EI}^{NL} w^{(3)} - k_{TM}^{NL} \right) - (-k_{EI} w^{(2)} -k_{2TM}) =\left. 0 \right|_{x=0} \\
\left( k_{ES}^{NL} u_0^{(2)} - k_{EI}^{NL} w^{(3)} - k_{TM}^{NL} \right) + (-k_{EI} w^{(2)} -k_{2TM}) =\left. 0 \right|_{x=L} \\
\end{array}
\end{equation}
where $\mathcal{N}_L, \mathcal{M}_L$ and $\mathcal{T}_L$ are unknown support reactions at $x=L$. The algebraic system of equations is solved for unknown constants and support reactions:
\begin{equation}\label{eq:Ex3_constants}
\begin{array}{c}
u_{00}= 0, u_1= -8.096,u_2= -16.57, u_3= -49.26, uw_{45}= 959.8 \\
w_0= 0, w_1= 0, w_2= 3.332, w_3= -52.77, w_4= 496.4 \\
\mathcal{N}_L = -5.196\cdot 10^{-5}, \mathcal{M}_L= -392.7, \mathcal{T}_L= -70.65.
\end{array}
\end{equation}
Final forms of displacement distributions can be finally written as:
\begin{equation}\label{eq:Ex3_u0}
\begin{array}{ll}
u_0(x)=&4.134 \cdot 10^{-4} x^3-6.202 \cdot 10^{-4} x^2+1.154 \cdot 10^{-4} x -3.034\\
&-0.268 e^{-9.534 x}-1.941 \cdot 10^{-5} e^{9.534 x}+2.084 e^{-5.141 x}\\
&+1.212\cdot 10^{-2} e^{5.141 x}+1.206 \cos (6.283 x)
\end{array}
\end{equation}
and
\begin{equation}\label{eq:Ex3_w}
\begin{array}{ll}
w(x)=&-2.584  \cdot 10^{-4} x^6+7.752  \cdot 10^{-4} x^5-1.460  \cdot 10^{-4} x^4+2.189 x^3\\
&-3.284 x^2+1.649 x -0.277\\
&+4.007 \cdot 10^{-2} e^{-9.534 x}-2.902 \cdot 10^{-6} e^{9.534 x}+0.238 e^{-5.141 x}\\
&-1.394 \cdot 10^{-3} e^{5.141 x}-5.545\cdot 10^{-3} \sin (6.283 x).\\
\end{array}
\end{equation}
Displacements are shown in Figs. \ref{fig:Ex3_u0} and \ref{fig:Ex3_w}. As expected, the U-shaped curve typical for transverse displacements of homogeneous local beams is not obtained. This can be attributed to complex interaction of the loading, coupling terms and layer distribution of the non-local beam. Strains arising from both mechanical and thermal loads, Fig. \ref{fig:Ex3_Eps} show a similar order of magnitude. Discontinuities are absent if the cumulative strain is considered, but not if only mechanical or only thermal strain is concerned. Thus, combination of two discontinuous fields gives a continuous one. In the case of stress field, discontinuities are also present in the total stress field. Higher stresses are obtained in the thinner layer.

\begin{figure}
	\centering
	\includegraphics[scale=0.42]{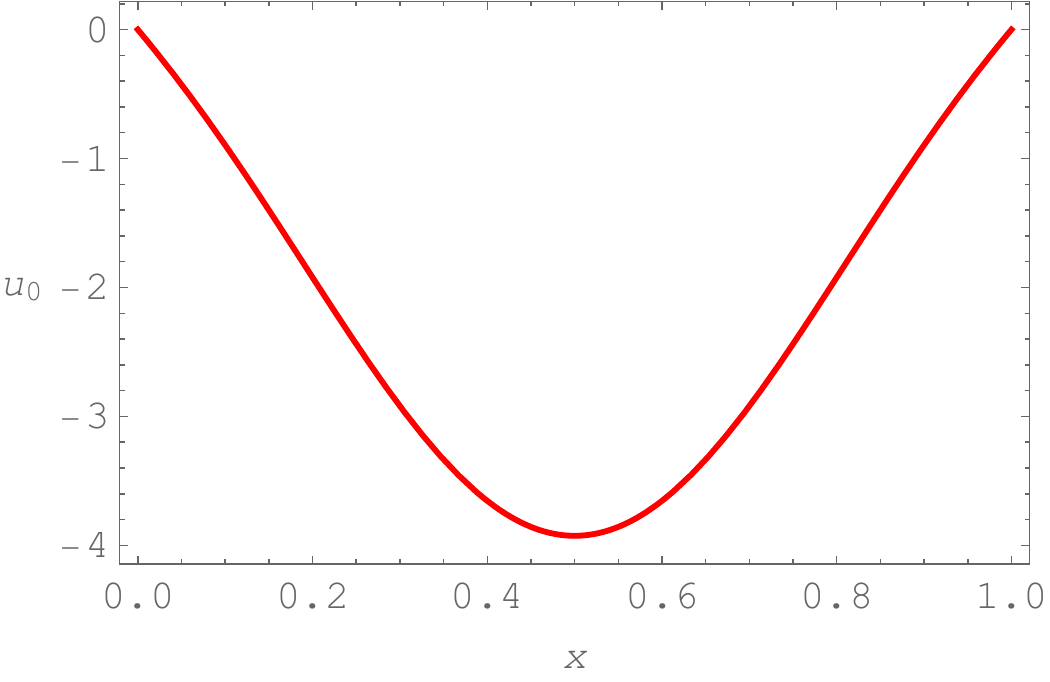}
	\caption{Axial displacements of the doubly clamped beam.} 
	\label{fig:Ex3_u0}
\end{figure}

\begin{figure}
	\centering
	\includegraphics[scale=0.42]{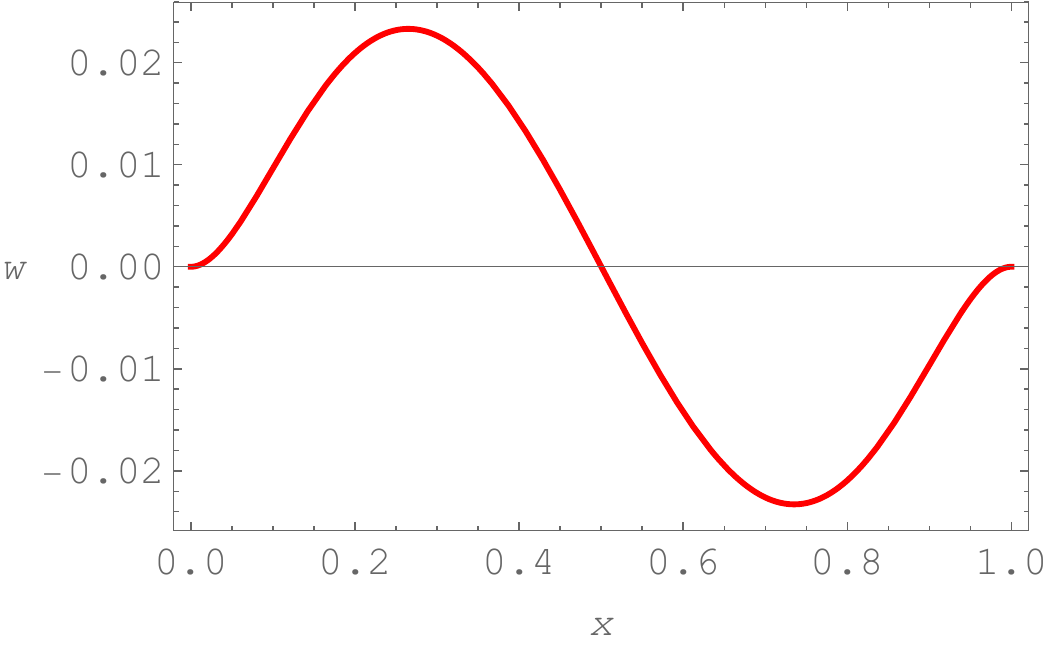}
	\caption{Transversal displacements of the doubly clamped beam.}
	\label{fig:Ex3_w}
\end{figure}

\begin{figure}
	\centering
	\includegraphics[scale=0.33]{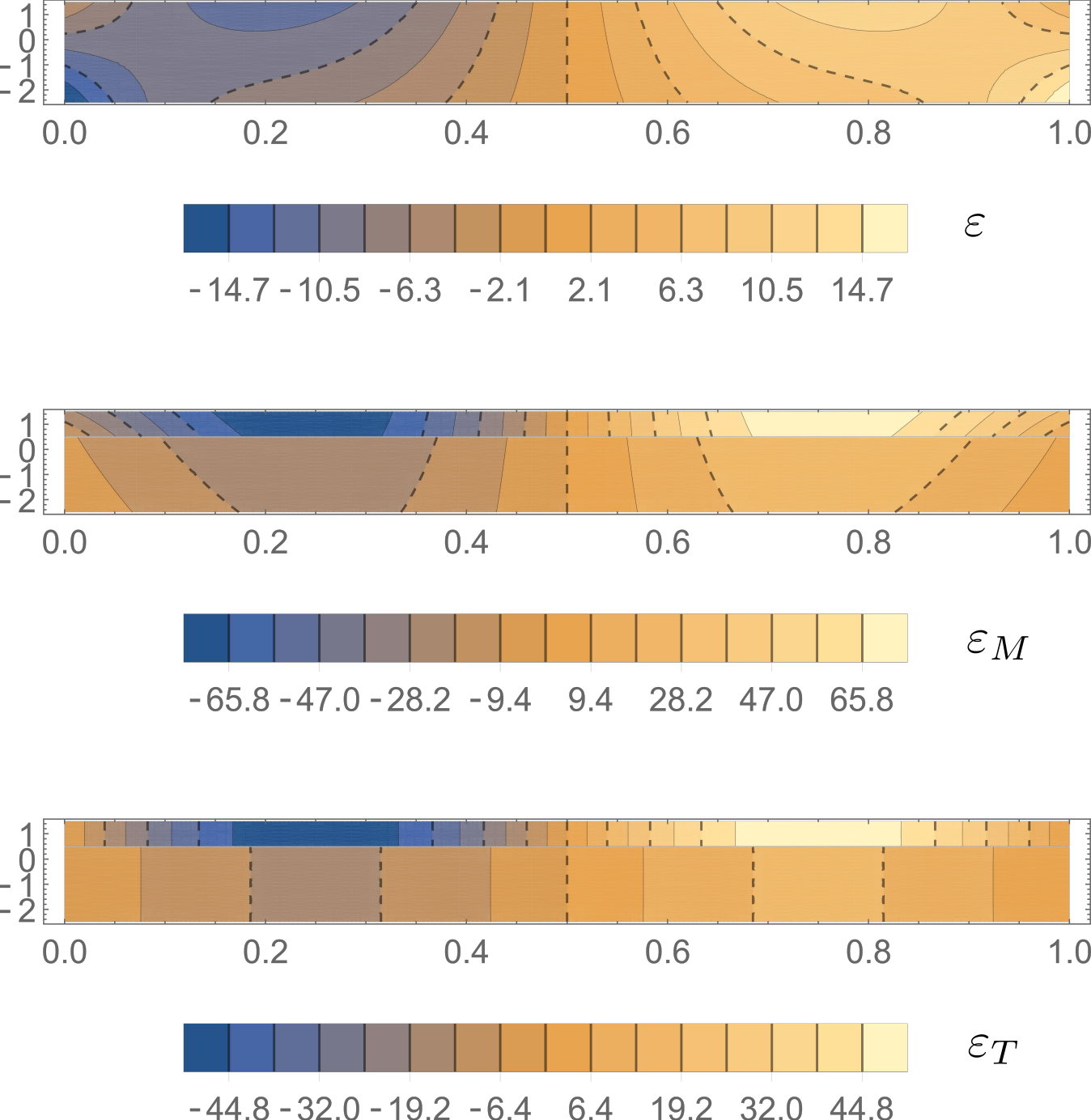}
	\caption{Strain distribution in the doubly clamped beam.} 
	\label{fig:Ex3_Eps}
\end{figure}

\begin{figure}
	\centering
	\includegraphics[scale=0.33]{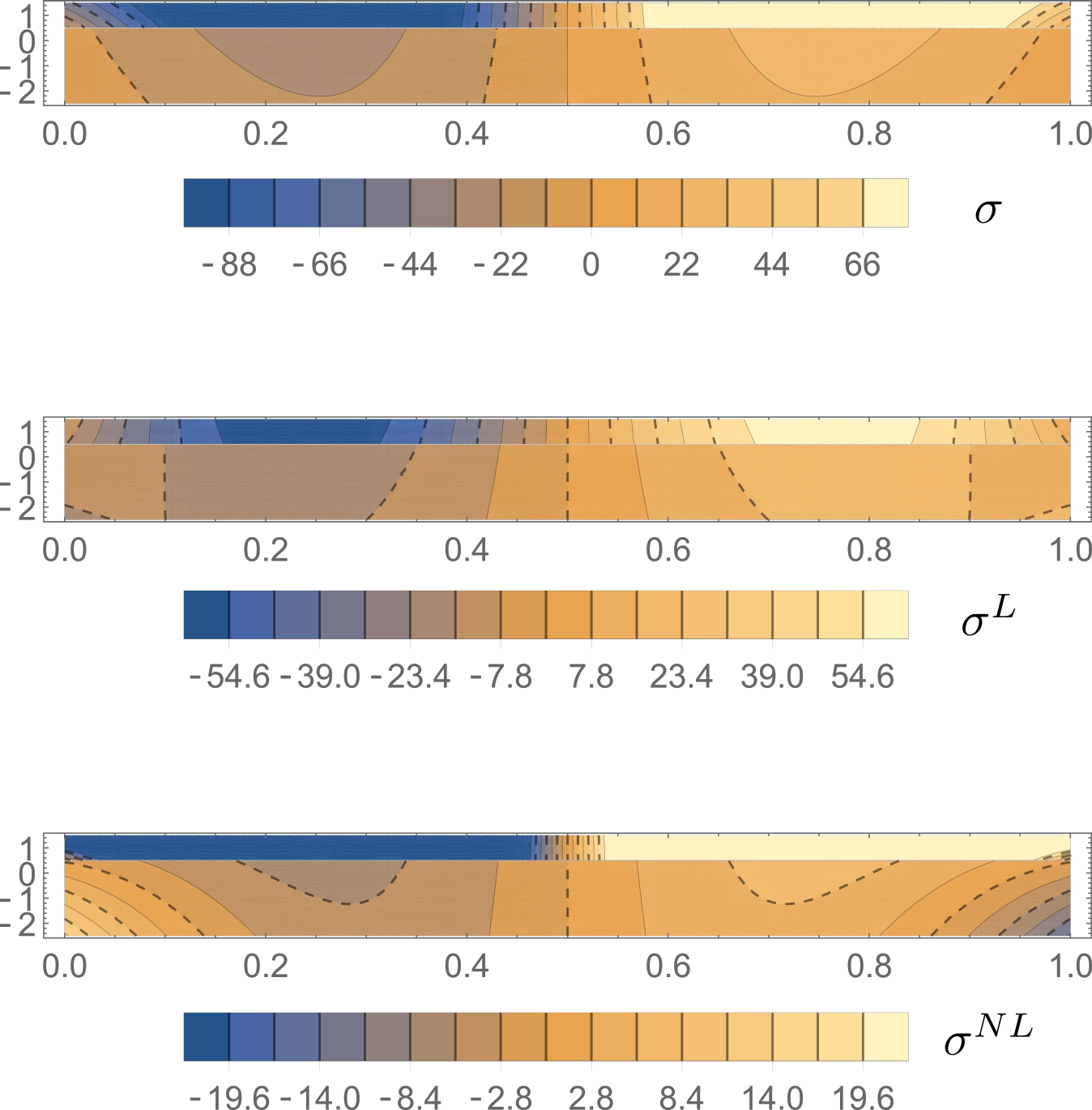}
	\caption{Total stress $\sigma=\sigma^L+\sigma^{NL}$, local part $\sigma^L$ of the stress and non-local part of the stress $\sigma^{NL}$ in a doubly clamped beam.} 
	\label{fig:Ex3_stress}
\end{figure}

\subsection{Influence of distribution of layers}
To evaluate different possibilities of layers' arrangement and their influence on mechanical behavior, a simply supported beam made of 4 layers is considered in the final example. Two different materials are used for layers and four different layered structures were considered. These are described by four-digits codes as 1122, 1221, 1212, 2112, where each digit denotes material ID, either 1 or 2, see Fig. \ref{fig:Ex4_beam}. Material data is again considered in non-dimensional form as $E_1=100, E_2=200$, $\alpha_1=0.1, \alpha_2=0.2$, $\lambda_1=0.1, \lambda_2=0.2$. The beam's width is constant $b=3$, while the height of each layer is $h_i=1$, where $i=1,2,3,4$. In that way the total height of the beam is $h=\sum_i h_i=4$. The beam's length is $L=100$. Beam is mechanically loaded with the distributed loading in both longituidnal $q_x=1$ and transverse direction $q_z=1$. As for the thermal loading, two cases are considered, Fig. \ref{fig:Ex4_beam}: 
\begin{enumerate}[label={(\alph*)}]
	\item  $\Delta \theta (x)=10 \frac{x}{L}$ and
	\item  $\Delta \theta (x,z)=10 \frac{x}{L} + 10 \frac{z}{h/2}$.
\end{enumerate}

\begin{figure}
	\centering
	\includegraphics[scale=0.42]{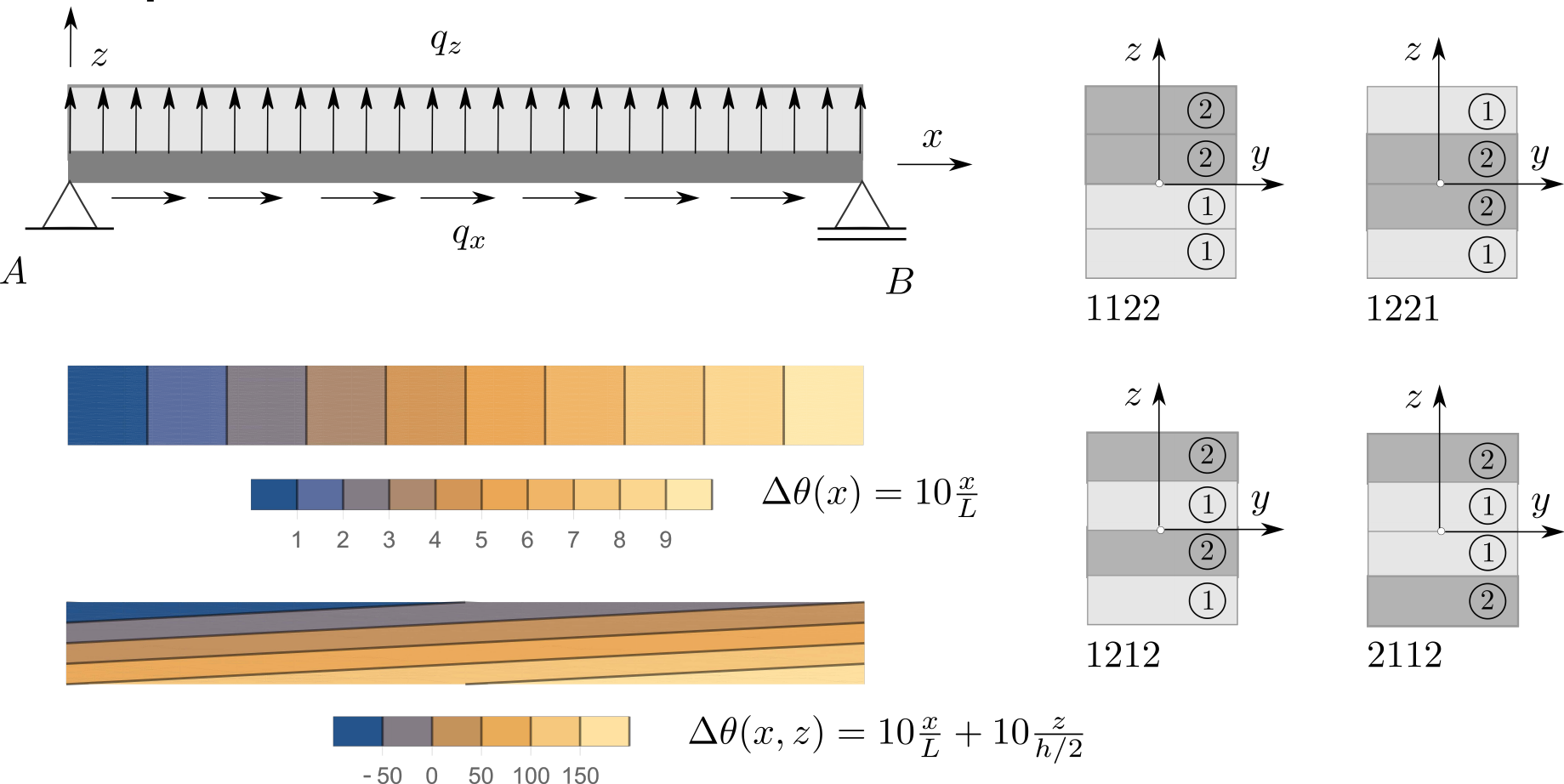}
	\caption{Simply supported beam and considered layered structures.} 
	\label{fig:Ex4_beam}
\end{figure}

The system of ordinary differential equations Eq.~(\ref{eq:Laplace}) is solved by Laplace transforms as before, with constitutive boundary conditions Eqs.~(\ref{eq:CBC}) and  boundary conditions Eqs.~(\ref{eq:BC}) as $u_{0}(0)=w(0)=w(L)=0$ and 
\begin{equation}\label{eq:Ex4_BC}
\begin{array}{ll}
\left. (-k_{EA}^{2NL} u_0^{(3)} +k_{ES}^{2NL}w ^{(4)} + k_{EA} u_0^{(1)} +k_{TN}) \right|_{x=L} = 0 \\
\left. (-k_{ES}^{2NL} u_0^{(3)} +k_{EI}^{2NL}w ^{(4)} -k_{EI} w ^{(2)}+k_{TM}) \right|_{x=0} = 0  \\
\left. (-k_{ES}^{2NL} u_0^{(3)} +k_{EI}^{2NL}w ^{(4)} -k_{EI} w ^{(2)}+k_{TM}) \right|_{x=L} = 0 \\ 
\end{array}
\end{equation}

Solutions are graphically presented in Figs. \ref{fig:Ex4_u0_a} - \ref{fig:Ex4_w_b}. Symmetric layer distributions 1221 and 2112 lead toward vanishing of the shift of the neutral surface. In these two cases terms $k_{ES}^{NL}=0$ and $k_{ES}^{2NL}=0$ so the problem is decoupled. In the Case (a), temperature does not change in the transverse direction. Nevertheless, the influence of different layer distributions can be noticed. Albeit the influence is small in the case of axial displacements, Fig. \ref{fig:Ex4_u0_a}, transverse displacements are markedly different, Fig. \ref{fig:Ex4_w_a}. Since the temperature does not change along the transverse coordinate, such behavior is mainly governed by the coupling effects.

In the Case (b), transverse temperature distribution also changes, so different layering schemes also affect bending behavior due to different coefficients of thermal expansions. This even causes minor loss of symmetry in transverse displacement distribution for the 1221 layer arrangement, Fig. \ref{fig:Ex4_w_b}. Difference is clearly visible in axial displacements, but almost identical solutions are obtained for symmetric schemes 1221 and 2112 (cannot be distinguished in Fig. \ref{fig:Ex4_u0_b}).

\begin{figure}
	\centering
	\includegraphics[scale=0.42]{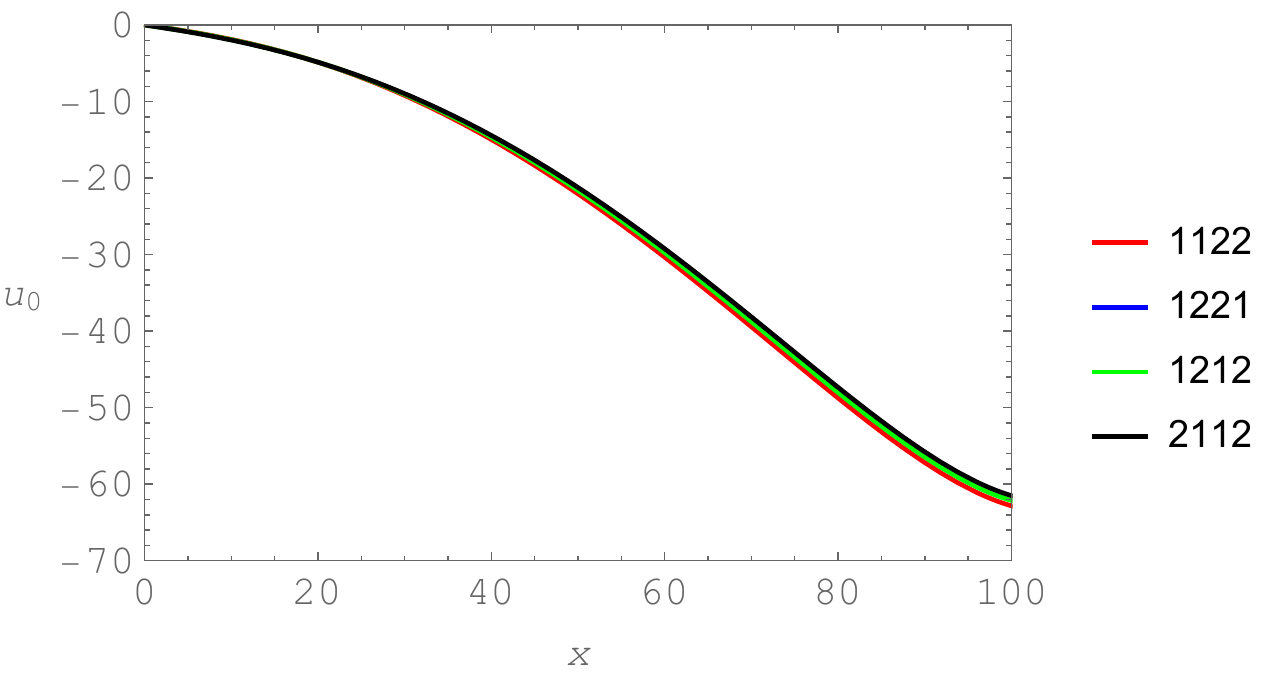}
	\caption{Axial displacements of the simply supported beam, Case (a).} 
	\label{fig:Ex4_u0_a}
\end{figure}

\begin{figure}
	\centering
	\includegraphics[scale=0.42]{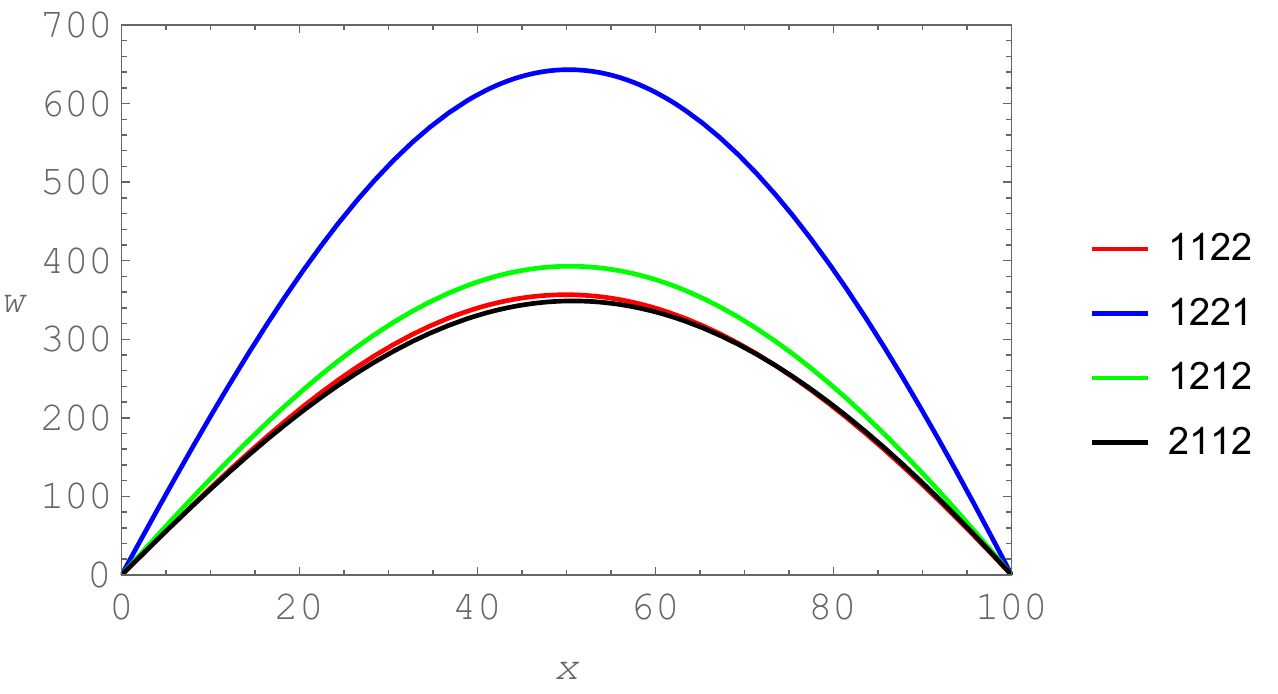}
	\caption{Transversal displacements of the simply supported beam, Case (a).}
	\label{fig:Ex4_w_a}
\end{figure}

\begin{figure}
	\centering
	\includegraphics[scale=0.42]{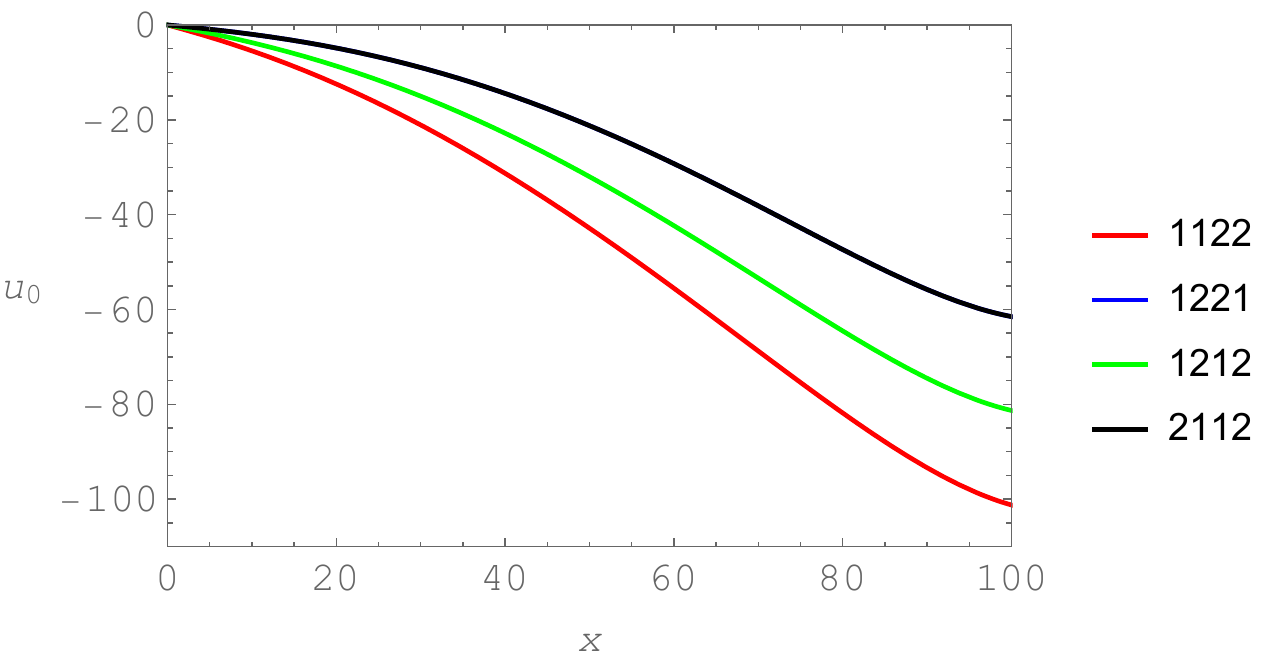}
	\caption{Axial displacements of the simply supported beam, Case (b).} 
	\label{fig:Ex4_u0_b}
\end{figure}

\begin{figure}
	\centering
	\includegraphics[scale=0.42]{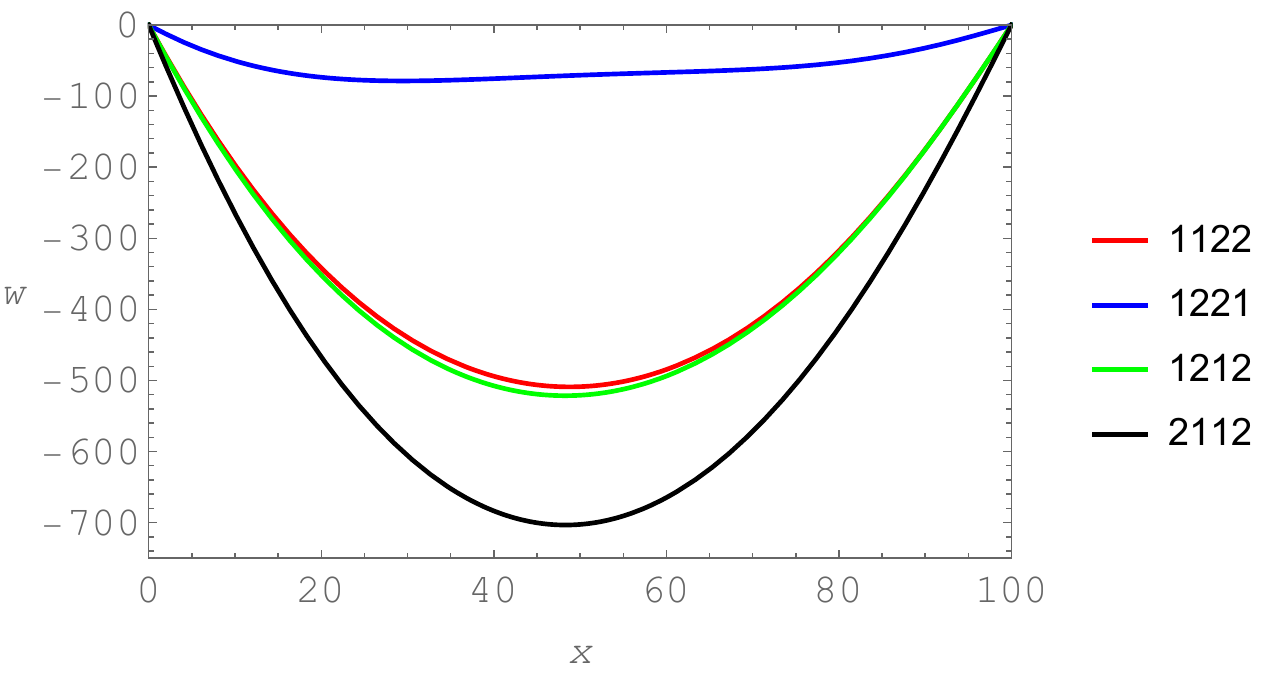}
	\caption{Transversal displacements of the simply supported beam, Case (b).}
	\label{fig:Ex4_w_b}
\end{figure}

\section{Closing remarks}
Thermomechanical analyses of multilayered micro- and nano-beams have been performed in this paper by the stress-driven nonlocal integral approach of elasticity. Main contributions and findings are summarized as follows.
\begin{itemize}
	\item Nonisothermal extension of existing isothermal nonlocal beam stress-driven integral formulations available in literature has been provided. The conceived methodology is suitable to model and assess the nonlocal behaviour of multilayered beams assembled of arbitrary number of layers made of different rectangular cross-sections. The proposed nonlocal strategy has been shown to be able to accomodate any kind of mechanical and thermal loading.
	\item Presence of terms that couple axial and transverse displacements is highlighted, due to the nonlocal nature of the presented model and occur both in isothermal and nonisothermal cases. Coupling contributions can be advantageously exploited to obtain some interesting mechanical responses with potential applications in thermomechanics of nanocomposites and new-generation Micro-/Nano-Electro-Mechanical-Systems (MEMS/NEMS). 
	\item These conclusions are supported by the in-depth analysis and numerical evidence provided in section 6, where selected case-studies of applicative interest in Nanoscience and Mechanics of Composites are investigated and commented upon. The illustrated examples can also serve as suitable benchmarks for further developments in thermomechanics of composite nanobeams.
\end{itemize}

\section*{Acknowledgments}
%This work has been supported in part by Croatian Science Foundation under the project IP-2019-04-4703 and by the the Italian Ministry for University and Research (P.R.I.N. National Grant 2017, Project code 2017J4EAYB; University of Naples Federico II Research Unit). This support is gratefully acknowledged.
This work has been fully supported by Croatian Science Foundation under the project IP-2019-04-4703. This support is gratefully acknowledged.

\bigskip

\noindent

\bibliographystyle{elsarticle-num}

\bibliography{SmallSizeParameter}

\end{document}